\documentclass[fleqn,12pt]{article}
\usepackage{amsfonts,amssymb,epsfig,latexsym,epic}
\usepackage[nospace]{cite}

\def\baselinestretch{1.1}
\makeatletter
\newbox\slashbox \setbox\slashbox=\hbox{$/$}
\def\pFMslash#1{\setbox\@tempboxa=\hbox{$#1$}
  \@tempdima=0.5\wd\slashbox \advance\@tempdima 0.5\wd\@tempboxa
  \copy\slashbox \kern-\@tempdima \box\@tempboxa}

\makeatother

\newcommand{\Gd}{\delta}
\newcommand{\Ge}{\epsilon}
\newcommand{\Geps}{\varepsilon}
\newcommand{\Gg}{\gamma}
\newcommand{\GG}{\Gamma}

\newcommand{\Gs}{\sigma}

\newcommand{\GT}{{\cal T}}

\newcommand{\GTh}{\Theta}

\newcommand{\Gz}{\zeta}

\newcommand{\cM}{{\scriptscriptstyle\cal M}}
\newcommand{\cN}{{\scriptscriptstyle\cal N}}
\newcommand{\cK}{{\scriptscriptstyle\cal K}}
\newcommand{\cL}{{\scriptscriptstyle\cal L}}
\newcommand{\cP}{{\scriptscriptstyle\cal P}}

\newcommand{\CA}{{\cal A}}

\newcommand{\CC}{{\cal C}}

\newcommand{\CI}{{\cal I}}
\newcommand{\CJ}{{\cal J}}
\newcommand{\CH}{{\cal H}}
\newcommand{\CM}{{\cal M}}
\newcommand{\CN}{{\cal N}}
\newcommand{\CK}{{\cal K}}
\newcommand{\CL}{{\cal L}}

\newcommand{\CP}{{\cal P}}
\newcommand{\CQ}{{\cal Q}}
\newcommand{\CS}{{\cal S}}
\newcommand{\CT}{{\cal T}}
\newcommand{\CV}{{\cal V}}

\newcommand{\dA}{{\dot{A}}}

\newcommand{\TCL}{{\tilde{\cal{L}}}}
\newcommand{\TCP}{{\tilde{\cal{P}}}}
\newcommand{\TCQ}{{\tilde{\cal{Q}}}}
\newcommand{\TCS}{{\tilde{\cal{S}}}}
\newcommand{\TCV}{{\tilde{\cal{V}}}}

\newcommand{\Ti}{{\tilde \imath}}

\newcommand{\Tj}{{\tilde \jmath}}
\newcommand{\Tr}{{\tilde r}}
\newcommand{\Ts}{{\tilde s}}

\newcommand{\Hi}{{\hat{\imath}}}
\newcommand{\Hj}{{\hat{\jmath}}}
\newcommand{\Hk}{{\hat{k}}}
\newcommand{\Hl}{{\hat{l}}}

\newcommand{\ft}[2]{{\textstyle {\frac{#1}{#2}} }}
\newcommand{\dd}{\partial}
\newcommand{\tr}{{\rm tr \,}}
\newcommand{\diag}{{\rm diag \,}}

\newcommand{\I}{{\rm i}}

\newcommand{\be}{\begin{equation}}
\newcommand{\ee}{\end{equation}}
\newcommand{\ben}{\begin{displaymath}}
\newcommand{\een}{\end{displaymath}}
\newcommand{\bea}{\begin{eqnarray}}
\newcommand{\eea}{\end{eqnarray}}
\newcommand{\ba}{\begin{eqnarray}}
\newcommand{\ea}{\end{eqnarray}}
\newcommand{\nn}{\nonumber}
\newcommand{\non}{\nonumber\\}
\newcommand{\bean}{\begin{eqnarray*}}
\newcommand{\eean}{\end{eqnarray*}}

\newcommand{\mathon}{\mathversion{bold}}
\newcommand{\mathoff}{\mathversion{normal}}

\def\moth{\mathsurround=0pt}
\newdimen\zo \zo=0pt

\def\tick{\leaders\hrule height 0.5ex depth 0pt \hskip 0.5pt}
\def\upboxfill{$\moth \setbox\zo\hbox{\tick}%
  \hskip 2pt\hbox to 0pt{$\tick$\hss}\hrulefill \hbox to
6pt{$\tick$\hss}$}

\def\dtick{\leaders\hrule height .34pt depth .5ex \hskip 0.5pt}
\def\downboxfill{$\moth \setbox\zo\hbox{\dtick}%
  \hskip 2pt\hbox to 0pt{$\dtick$\hss}\hrulefill \hbox to
6pt{$\dtick$\hss}$}

\newcommand{\la}{\label}

\newcommand{\Ref}[1]{(\ref{#1})}

\makeatletter
\@addtoreset{equation}{section}
\makeatother

\newcommand{\SU}[1]{{{\rm SU}({#1})}}
\newcommand{\SO}[1]{{{\rm SO}({#1})}}

\newcommand{\vl}{{\vphantom{[}}}

\newcommand{\cro}{\!\times\!}

\newcommand{\pls}{\!+\!}
\newcommand{\mis}{\!-\!}

\begin{document}

\thispagestyle{empty}

\begin{flushright}
AEI-2003-051 \\
ITP-UU-03/32 \\
SPIN-03/20
\end{flushright}

\mathon
\begin{center}
{\bf\Large Kaluza-Klein supergravity on AdS$_3\times S^3$}
\mathoff

\mathoff

\bigskip\bigskip

{{\bf Hermann~Nicolai\footnote{\tt nicolai@aei.mpg.de} and
Henning~Samtleben\footnote{\tt h.samtleben@phys.uu.nl}}}

\vspace{.3cm}
$^1${\small Max-Planck-Institut f{\"u}r Gravitationsphysik\\
  Albert-Einstein-Institut\\
  M\"uhlenberg 1, D-14476 Potsdam, Germany}

\vspace{.5cm}

$^2${\small Institute for Theoretical Physics \& Spinoza Institute\\
Utrecht University, Postbus 80.195 \\
3508 TD Utrecht, The Netherlands}

\end{center}
\renewcommand{\thefootnote}{\arabic{footnote}}
\setcounter{footnote}{0}
\bigskip
\medskip
\mathoff
\begin{abstract}
We construct a Chern-Simons type gauged $N=8$ supergravity in three
spacetime dimensions with gauge group $\SO4\times T_\infty$ over the 
infinite dimensional 
coset space $\SO{8,\infty}/\left(\SO8\times \SO{\infty}\right)$, where 
$T_\infty$ is an infinite dimensional translation subgroup of $\SO{8,\infty}$. 
This theory describes the effective interactions of the (infinitely many) 
supermultiplets contained in the two spin-1 Kaluza-Klein towers arising 
in the compactification of $N=(2,0)$ supergravity in six dimensions 
on AdS$_3 \times S^3$ with the massless supergravity multiplet. After
the elimination of the gauge fields associated with $T_\infty$, one
is left with a Yang Mills type gauged supergravity with gauge group
$SO(4)$, and in the vacuum the symmetry is broken to the (super-)isometry 
group of AdS$_3 \times S^3$, with infinitely many fields acquiring masses 
by a variant of the Brout-Englert-Higgs effect.
\end{abstract}

\renewcommand{\thefootnote}{\arabic{footnote}}
\setcounter{footnote}{0}

\newpage

\mathoff
\section{Introduction}

One of the best studied examples of the celebrated AdS/CFT duality
conjecture~\cite{Mald97,AGMOO00} is the D1-D5 system, which relates
IIB string theory on AdS$_3\times S^3\times M_4$ to a two-dimensional
$N=(4,4)$ supersymmetric conformal field theory (CFT) living on the
boundary of AdS$_3$~\cite{MalStr98,Mart98,GiKuSe98,SeiWit99,DaMaWa02,Giveon:2003ku}.
The latter is believed to be described by a non-linear $\Gs$-model
whose target space is a deformation of the symmetric orbifold
$(M_4)^n/S_n$~\cite{Vafa95,Witt97,LarMar99}. In the supergravity
limit, the Kaluza-Klein (KK) modes on the AdS$_3\times S^3\times M_4$
near horizon geometry of the D1-D5 system are dual to chiral primary
operators in the conformal field theory. Although CFT calculations
have been mainly performed at the `orbifold
point'~\cite{JeMiRa99,LunMat01}, where the supergravity approximation
breaks down, nontrivial tests of the correspondence are possible for
quantities protected by non-renormalization theorems; in particular,
BPS spectra and elliptic genera were matched successfully
\cite{Lars98,dBoe98a,MaMoSt99,GHMN00}.

The computation of higher point correlation functions requires the
evaluation of higher order supergravity couplings which have been
extensively studied for the compactification of IIB supergravity on
AdS$_5\times S^5$~\cite{FMMR98,LMRS98,Lee99,AruFro99} and for
AdS$_3\times S^3$ \cite{Miha99,ArPaTh00}. With increasing order of
interactions, however, this program becomes stymied by a plethora of
non-linear field redefinitions. Nevertheless, for the AdS$_5\times
S^5$ compactification, the effective low energy theory for the lowest
(massless) supermultiplet at the bottom of the KK tower is believed to
be known to {\em all} orders and to coincide with the maximal $D=5$,
$N=8$ gauged supergravity with gauge group ${\rm SO}(6)$
\cite{GuRoWa86}.\footnote{To be sure, the consistency of this
truncation has never been fully established despite much supporting
evidence (see e.g. \cite{PilWar00} and references therein), unlike for
the AdS$_4 \times S^7$ \cite{deWNic87} and AdS$_7 \times S^4$
\cite{NaVavN99} truncations of $D\!=\!11$ supergravity.} In
particular, the scalar potential of this theory carries information
about the deformations of the dual CFT by relevant operators and the
corresponding renormalization group
flows~\cite{FGPW99,GPPZ00,BiFrSk01}. It would clearly be desirable to
have an effective theory describing the full non-linear couplings of
the higher KK supermultiplets as well, but this does not appear
possible in five dimensions due to the impossibility of consistently
coupling a finite number of massive spin-2 fields.  In three
dimensions, the situation is different. The AdS$_3\times S^3$
background is only half maximally supersymmetric, and instead of a
single tower of supermultiplets as for AdS$_5\times S^5$, there are
{\em three} different KK towers in the reduction.  One of these
contains the massive spin-2 supermultiplets (with the massless $N=8$
supergravity multiplet at the bottom), while the other two consist of
spin-1 supermultiplets~\cite{dBoe98a,DKSS98}.  In view of the duality
between vector gauge fields and scalar fields in three dimensions it
is therefore plausible that there should exist a unified description
at least of the two spin-1 towers in terms of an infinite number of
$N=8$ supermultiplets coupled to the massless (nonpropagating) $N=8$
supergravity multiplet.

In this paper, we will demonstrate that such a construction is indeed 
possible, and present an effective three-dimensional theory that
describes the massless $N=8$ supergravity multiplet and the entire 
two infinite spin-1 towers and their interactions in terms of a gauged 
supergravity over a {\em single} irreducible coset space. Furthermore, 
we will show that the spin-1 towers can be consistently truncated to 
any finite subset of spin-1 multiplets. Our construction exploits the 
special properties of gauged supergravities in three 
dimensions \cite{NicSam00,NicSam01b,LuPoSe02,NicSam03a,LuPoSe03,
dWHeSa03,FiNiSa03}, which have no analog in dimensions $D\geq 4$, 
and makes essential use of the results of~\cite{NicSam03a} 
establishing the link between Yang-Mills (YM) and Chern-Simons (CS) 
type gauged supergravities in three dimensions. It follows from these 
results that all the relevant information about the effective 
$D=3$ theory is encoded in the infinite-dimensional coset space 
${\rm G}/{\rm H}=\SO{8,\infty}/ (\SO8\times\SO\infty)$, or more precisely,
\ba
{\rm G}/{\rm H} &=&
{\rm SO}\Big(8\;,\,{ \sum_{k\ge2} k^2} +
n {\sum_{l\ge1} l^2} \Big) \Big/
\Big( \SO8\times
{\rm SO}\Big({ \sum_{k\ge 2} k^2} +
n {\sum_{l\ge1} l^2} \Big) \Big)
\;.
\la{GI}
\ea
The parameter $n$ denotes the number of tensor multiplets in the
six-dimensional theory which is $n=5$, and $n=21$ for $M_4=T^4$ and
$M_4=K_3$, respectively. Instead of the infinite sums, one may for
definiteness consider any finite truncation which yields a consistent
and supersymmetric theory, coupling a finite but arbitrarily large
number of supermultiplets to the basic $N=8$ supergravity
Lagrangian. In particular, we give the scalar potential as a function
on the coset manifold \Ref{GI} which yields the KK scalar and vector
masses by virtue of a three-dimensional variant of the
Brout-Englert-Higgs mechanism on the infinite-dimensional
space~\Ref{GI}. The fact that this agreement extends to the complete
self-interactions induced by the KK compactification is a consequence
of the uniqueness of the effective locally $N=8$ supersymmetric
theory. The extension of these results to the spin-2 tower, and hence
to the {\em full} KK theory, remains an open problem for the time
being; see, however, the comments in section~5. We emphasize that we
do not wish to address here the issue of consistency of the KK
truncation from six dimensions, but at this stage focus on the
consistent three-dimensional theory. \footnote{So far, consistency of
truncations has been shown for the lowest supermultiplets in the
$N=(1,0)$ six-dimensional theory by explicitly constructing the
non-linear ansatz in the higher-dimensional
theory~\cite{LuPoSe02,LuPoSe03}. An order by order analysis of the
consistency for the higher modes was initiated in \cite{ArPaTh00}.}
Settling this issue will presumably require the inclusion of the
spin-2 tower into the analysis.

We note that, already some time ago and in a different context, the
idea of describing the effective interactions of an infinite number of
fields in terms of a gauged supergravity was proposed in an attempt to
describe the effective interactions of the massive scalar string modes
arising in the compactification to four dimensions in terms of an
$N=4$ gauged supergravity over the coset ${\rm SO}(6,\infty)/({\rm
SO}(6) \times {\rm SO}(\infty))$ \cite{GivPor90}.  There the relevant
gauge group, which must be a subgroup of ${\rm SO}(6,\infty)$, is
based on an indefinite lattice algebra of string vertex operators.

The paper is organized as follows. In section~2 we briefly review the
KK spectrum of six-dimensional $N=(2,0)$ supergravity on AdS$_3\times
S^3$. In particular, we discuss the lowest floors of the three KK
towers, given by the massless (nonpropagating) supergravity multiplet,
a short spin-$\ft12$ matter multiplet, and the massive spin-$1$
multiplet containing the YM vector fields. In section~3, we present
the effective three-dimensional theory which describes the coupling of
these three multiplets alternatively as an $\SO4$ YM theory, or as a
CS theory with gauge group $\SO4\ltimes {\rm T}_6$. The theory is
extended in section~4 to include massive spin-$1$ multiplets of
arbitrarily high KK level. The construction is based on the
infinite-dimensional coset space~\Ref{GI} while the CS gauge group is
enlarged to $\SO4\ltimes {\rm T}_\infty$, where ${\rm T}_\infty$
denotes an infinite translational subgroup of ${\rm H}$. We close in
section~5 with a few comments on the possible inclusion of the spin-2
tower.

\mathon
\section{Spectrum of supergravity on AdS$_3 \times S^3$}
\mathoff

The mass spectrum of six-dimensional $N=(2,0)$ supergravity on
AdS$_3\times S^3$ has been computed in~\cite{DKSS98} by linearizing 
the equations of motion around the AdS background, and in~\cite{dBoe98a} 
by group theoretical arguments in terms of unitary irreducible representations
of the supergroup $\SU{2|1,1}_L\times\SU{2|1,1}_R$. Here, we briefly
review these results to the extent needed in the following. The
field content of the six-dimensional theory~\cite{Roma86}
comprises the supergravity multiplet with graviton, gravitini and
five self-dual tensor fields, and $n$ tensor multiplets, each
containing an anti-selfdual tensor field, four fermions and five
scalars. The scalar sector forms a coset space $\sigma$-model
$\SO{5,n}/(\SO{5}\cro\SO{n})$. The AdS$_3\times S^3$ background
endows one of the five tensor fields of the supergravity
multiplet with a vacuum expectation value
\ba\label{FreundRubin} B^5_{\mu\nu\rho} = f \Geps_{\mu\nu\rho}
\;\; , \quad B^5_{mnp} = f \Geps_{mnp}\;,\qquad B^\Ti_{\mu\nu\rho}
=0=B^\Ti_{mnp}\;,\quad \Ti=1, \dots, 4 \;, \label{FR} \ea
where $f$ is the Freund-Rubin parameter. For $f\neq 0$, the
$R$-symmetry group is broken from $\SO{5}$ down to $\SO{4}$.
Together with the $\SO{n}$ rotating the tensor multiplets, this
group survives as a global symmetry of the three-dimensional
effective theory.

The spectrum of the three-dimensional theory is hence organized under
the AdS$_3$ supergroup $\SU{2|1,1}_L\times\SU{2|1,1}_R$ whose bosonic
extension $\SO{3}_L\times\SO{3}_R\equiv \SO{4}_{\rm gauge}$
corresponds to the isometry group of the three-sphere $S^3$, and a
global $\SO{4}\times\SO{n}$. It consists of three Kaluza-Klein (KK)
towers: a spin-2 tower, and two spin-1 towers transforming as vector
and singlet under $\SO{n}$, respectively.  For later use, we give the
generic spin-1 multiplet in table~\ref{spin1}. It contains $16k^2$
degrees of freedom, and following~\cite{dBoe98a} we will designate it
by $({\bf k\pls1},{\bf k\pls1})_S$. The $\SO4$ representations are
labeled by their spins $[j_1,j_2]$ under $\SO3_L\times\SO3_R$ while the
numbers $(\Delta,s_0)$ label the representations of the AdS group
$\SO{2,2}$.

\begin{table}[bt]
{\small
\begin{center}
\begin{tabular}{crcccc}\hline
$\Delta$ & $s_0$ & & ${\rm SO}(4)_{\rm gauge} $ &
${\rm SO}(4)_{\rm glob}$ &
\# dof
\\ \hline\hline
$k$ & $0$&  & $\left[\ft{k}2,\ft{k}2\right]$ & $\left[0,0\right]$ &
$(k+1)^2$  \\
$k+\ft12$ &$\ft12$& & $\left[\ft{k}2,\ft{k-1}2\right]$ &
$\left[0,\ft12\right]$  &
$2k(k+1)$
\\
$k+\ft12$&$-\ft12$&  & $\left[\ft{k-1}2,\ft{k}2\right]$ &
$\left[\ft12,0\right]$  &
$2k(k+1)$
\\
$k+1$ &
$0$&&$\left[\ft{k-1}2,\ft{k-1}2\right]$&$\left[\ft12,\ft12\right]$  &
$4k^2$
\\
$k+1$ & $1$ && $\left[\ft{k}2,\ft{k-2}2\right]$ & $\left[0,0\right]$
& $k^2-1$  \\
$k+1$ & $-1$ & & $\left[\ft{k-2}2,\ft{k}2\right]$ & $\left[0,0\right]$
& $k^2-1$ \\
$k+\ft32$ & $\ft12$&  & $\left[\ft{k-1}2,\ft{k-2}2\right]$ &
$\left[\ft12,0\right]$  &  $2k(k-1)$
\\
$k+\ft32$ & $-\ft12$ & & $\left[\ft{k-2}2,\ft{k-1}2\right]$ &
$\left[0,\ft12\right]$  &  $2k(k-1)$
\\
$k+2$ &  $0$ & & $\left[\ft{k-2}2,\ft{k-2}2\right]$  &
$\left[0,0\right]$ & $(k-1)^2$  \\
\hline
\end{tabular}
\caption{Spin-1 multiplet $({\bf k\pls1},{\bf k\pls1})_S$ of
$\SU{2|1,1}_L\times\SU{2|1,1}_R$.}
\la{spin1}
\end{center}}
\end{table}

Let us describe in a little more detail the lowest levels of the three
KK towers which we have collected in table~\ref{spectrum}.  For
details on the higher dimensional origin of the modes, indicated in
the last column, we refer to \cite{DKSS98}. Recall that the
six-dimensional theory does not admit a Lagrangian formulation due to
the self-duality constraint obeyed by the two-forms $B_{MN}$. Upon
compactification on $S^3$, these self-duality equations are used to
eliminate the components $B_{\mu\nu}$ from the theory. The spin-2
tower starts from the massless supergravity multiplet, which in three
dimensions does not carry propagating degrees of freedom. It comprises
the metric, gravitinos and pure gauge modes of the $\SO{4}$ vectors,
see below. The lowest level of the spin-1 $\SO{n}$-vector tower is
occupied by the degenerate short (spin-$\ft12$) multiplet $({\bf
2},{\bf 2})_S$ of table~\ref{spin1} that contains 8 scalars and 8
fermions, all transforming in the vector representation of $\SO{n}$,
labeled by the index~$\Tr$. By contrast, the spin-1 $\SO{n}$-singlet
tower starts from the generic multiplet $({\bf 3},{\bf 3})_S$ of
table~\ref{spin1} whose bosonic content is given by 26 scalars and 6
propagating vector fields.

\begin{table}[bt]
{\small
\begin{center}
\begin{tabular}{crcccrc}\hline
$\Delta$ & $s_0$ & ${\rm SO}(4)_{\rm gauge} $ &
${\rm SO}(4)_{\rm glob}$ & ${\rm SO}(n)_{\rm glob}$ &
\# dof & 6d origin
\\ \hline\hline
\multicolumn{7}{c}{Nonpropagating gravity multiplet $({\bf
3,1})_S+({\bf 1,3})_S$}
\\
\hline
$2$ & $2$  & $\left[0,0\right]$ & $\left[0,0\right]$ & $1$ & $0$
& $g_{\mu\nu}$ \\
$2$ & $-2$  & $\left[0,0\right]$ & $\left[0,0\right]$ & $1$ & $0$
& $g_{\mu\nu}$ \\
$\ft32$& $\ft32$  & $\left[0,\ft12\right]$ & $\left[0,\ft12\right]$ &
$1$ & $0$
& $\psi_\mu$ \\
$\ft32$& $-\ft32$  & $\left[\ft12,0\right]$ & $\left[\ft12,0\right]$ &
$1$ & $0$
& $\psi_\mu$ \\
$1$ & $1$  & $\left[0,1\right]$ & $\left[0,0\right]$ & $1$ & $0$
& $g_{\mu m}, B^5_{\mu m}$ \\
$1$ & $-1$  & $\left[1,0\right]$ & $\left[0,0\right]$ & $1$ & $0$
& $g_{\mu m}, B^5_{\mu m}$ \\
\hline
\multicolumn{7}{c}{Spin-$\ft12$ hypermultiplet $({\bf 2,2})_S$} \\
\hline
$1$ & $0$  & $\left[\ft12,\ft12\right]$ & $\left[0,0\right]$ & $n$ & $4n$
& $\phi^{5\Tr}, B^\Tr_{mn}$ \\
$\ft32$&$\ft12$   & $\left[\ft12,0\right]$ & $\left[0,\ft12\right]$ &
$n$ & $4n$
& $\chi^\Tr$ \\
$\ft32$ & $-\ft12$ & $\left[0,\ft12\right]$ & $\left[\ft12,0\right]$ &
$n$ & $4n$
& $\chi^\Tr$ \\
$2$ & $0$ & $\left[0,0\right]$ & $\left[\ft12,\ft12\right]$ & $n$ &
$4n$  & $\phi^{\Ti\Tr}$  \\
\hline
\multicolumn{7}{c}{Spin-$1$ multiplet $({\bf 3,3})_S$} \\ \hline
$2$ & $0$  & $\left[1,1\right]$ & $\left[0,0\right]$ & $1$ & $9$
& $B^5_{mn}, g_m{}^m, g_\mu{}^\mu$ \\
$\ft52$ &$\ft12$ & $\left[1,\ft12\right]$ & $\left[0,\ft12\right]$ &
$1$ & $12$
& $\psi_m$  \\
$\ft52$&$-\ft12$  & $\left[\ft12,1\right]$ & $\left[\ft12,0\right]$ &
$1$ & $12$
& $\psi_m$   \\
$3$ & $0$&$\left[\ft12,\ft12\right]$&$\left[\ft12,\ft12\right]$ & $1$
& $16$
&$ B^{\Ti}_{mn}$ \\
$3$ & $1$ & $\left[1,0\right]$ & $\left[0,0\right]$ & $1$ & $3$
& $g_{\mu m}, B^5_{\mu m}$ \\
$3$ & $-1$  & $\left[0,1\right]$ & $\left[0,0\right]$ & $1$ & $3$
& $g_{\mu m}, B^5_{\mu m}$ \\
$\ft72$ & $\ft12$  & $\left[\ft12,0\right]$ & $\left[\ft12,0\right]$ &
$1$ & $4$
& $\psi_m$   \\
$\ft72$ & $-\ft12$  & $\left[0,\ft12\right]$ & $\left[0,\ft12\right]$
& $1$ & $4$
& $\psi_m$   \\
$4$ &  $0$  & $\left[0,0\right]$ & $\left[0,0\right]$ & $1$ & $1$
& $g_m{}^m, g_\mu{}^\mu$ \\
\hline
\end{tabular}
\caption{Lowest mass spectrum on $AdS_3 \times S^3$.}
\la{spectrum}
\end{center}}
\end{table}

The six-dimensional origin of the vector fields in these multiplets is
somewhat subtle due to the mixing between the KK vectors $g_{\mu m}$
and the tensor components $B^5_{\mu m}$.  Parametrizing the lowest
order fluctuations of the metric and the distinguished tensor
field~\Ref{FR} as $g_{\mu m}\equiv K_\mu^{\pm} Y_m^{(1,\pm1)},
B^5_{\mu m}\equiv Z_\mu^{\pm} Y_m^{(1,\pm1)}$ with the lowest $S^3$
vector harmonics $Y_m^{(1,\pm1)}$, one arrives at the following
linearized coupled system of YM and CS equations 
\ba
\nabla^\nu K^{\pm}_{\mu\nu} -
\frac4{L_0}\,\Ge_{\mu\nu\rho}Z^{\pm\,\nu\rho}
&=& 0
\;,\qquad
\Ge_{\mu\nu\rho}Z^{\pm\, \nu\rho} +
\frac2{L_0}\left(K^{\pm}_\mu \pm 2 Z^{\pm}_\mu\right) ~=~ 0 \;,
\la{vecfluc}
\ea
where $K_{\mu\nu}\equiv \partial_\mu K_\nu - \partial_\nu K_\mu$
and $Z_{\mu\nu}\equiv \partial_\mu Z_\nu - \partial_\nu Z_\mu$,
with remaining gauge freedom $\delta K_\mu^{\pm} = \dd_\mu
\Lambda^\pm$, $\delta Z_\mu^{\pm} = \mp\ft12 \dd_\mu
\Lambda^\pm$ and the AdS length $L_0$. The system \Ref{vecfluc}
has three eigenmodes
\ba
K^{\pm}_\mu = \mp 2 Z^{\pm}_\mu
\;,\qquad
K^{\pm}_\mu = \mp 4 Z^{\pm}_\mu
\;,\qquad
K^{\pm}_\mu = \pm 2 Z^{\pm}_\mu
\;.
\la{vecmodes}
\ea
The first mode preserves gauge invariance and leads to the pure gauge
states of the gravity multiplet in table~\ref{spectrum}. The second
mode in \Ref{vecmodes} yields the propagating vectors of the spin-$1$
multiplet in table~\ref{spectrum}, carrying one degree of freedom and
satisfying a massive CS equation $\Ge_{\mu\nu\rho}K^{\pm\nu\rho}=
\pm(4/L_0)K^\pm_\mu$. Alternatively, one may assemble this mode together
with the non-propagating gauge mode above into a vector field
satisfying the gauge covariant YM equation $\nabla^\nu K^\pm_{\mu\nu}
\pm (1/L_0)\Ge_{\mu\nu\rho}K^{\pm\,\nu\rho} =0$ with topological mass
term. Finally, the third eigenmode in \Ref{vecmodes} gives rise to
massive CS vectors that are located in the lowest massive multiplet of
the spin-$2$ tower~\cite{DKSS98}.

In the next section we will present a three-dimensional supergravity
theory with local $\SO4$ symmetry that combines all the lowest level
multiplets given in table~\ref{spectrum} and admits an $N=(4,4)$
supersymmetric AdS$_3$ groundstate. From the above discussion, we
expect the vector fields of this theory to be given by either $12$ CS
fields whose equations of motion linearized around the groundstate
take the form
\ba
\Ge_{\mu\nu\rho}K^{\pm\nu\rho} &=& \pm\frac4{L_0}\,K^\pm_\mu
\;,\qquad \mbox{and}\qquad
\Ge_{\mu\nu\rho}\hat{K}^{\pm\nu\rho} ~=~ 0 \;,
\qquad \mbox{respectively}\; ,
\la{vclin1}
\ea
or by a set of $6$ YM vector fields, satisfying equations
\ba
\nabla^\nu K^\pm_{\mu\nu} \pm \frac1{L_0}\,
\Ge_{\mu\nu\rho}K^{\pm\,\nu\rho}
&=& 0 \;.
\la{vclin2}
\ea
In \cite{NicSam03a} we have established the equivalence of YM
gaugings with CS gaugings in three dimensions, in the sense that a
YM gauged supergravity with gauge group $G_0$ is equivalent on
shell to a CS gauged supergravity with gauge group $G_0 \ltimes
T$, where $T$ is a translation group containing a subgroup
transforming in the adjoint of $G_0$.\,\footnote{Although fermionic 
terms were not explicitly considered in \cite{NicSam03a}, 
it is straightforward to determine the modifications of the 
supersymmetry variations and the YM type Lagrangian coming from 
the elimination of the translational CS vector fields by retaining 
the fermionic bilinears in their equations of motion. However, the
resulting YM type Lagrangian has many more terms than the original 
CS type Lagrangian, which makes the comparison with a direct construction 
of the YM type theory somewhat cumbersome.} In the following we shall 
exploit this result in order to construct a three-dimensional 
theory whose two equivalent formulations give rise to vector 
equations of the form \Ref{vclin1} and \Ref{vclin2}, respectively.

\section{The three-dimensional theory with YM multiplet}

In this section we present the construction of the three-dimensional
theory that describes the coupling of the lowest level multiplets of
the three KK towers, collected in table~\ref{spectrum}. The
supergravity multiplet contains the non-propagating fields in three
dimensions. The corresponding topological supergravity theory has been
given in~\cite{AchTow86,Davi99} as a Chern-Simons theory based on the
supergroup~$\SU{2|1,1}_L\times\SU{2|1,1}_R$. The coupling of this
theory to propagating matter in the $n$ spin-$\ft12$ hypermultiplets
$({\bf 2},{\bf 2})_S$ has been constructed in~\cite{NicSam01b}. It
comes as an $\SO{8,n}/(\SO{8}\times\SO{n})$ coset space model with
$\SO{4}$ gauge group and non-propagating CS vector fields; its scalar
potential has been studied in the context of holographic RG flows
in~\cite{BerSam01}.

We shall now extend this construction to include the coupling to the
spin-$1$, $\SO{n}$ singlet, supermultiplet $({\bf 3},{\bf 3})_S$ of
table~\ref{spectrum} which contains the $\SO{4}$ Yang-Mills gauge
vectors, in order to describe the full lowest mass spectrum of
supergravity on AdS$_3 \times S^3$. The construction follows the
strategy outlined in~\cite{NicSam03a} yielding a CS gauged
supergravity with the particular non-semisimple gauge group that
allows for an on shell equivalent formulation with propagating YM
gauge fields.

First of all, counting of degrees of freedom we see that after
dualizing all degrees of freedom into the scalar sector, the spectrum
of table~\ref{spectrum} consists of $32+8n$ bosonic and the same
number of fermionic degrees of freedom. The theory describing this
field content should thus be obtainable as a CS gauging of the $N=8$
theory with coset space
\ba
{\rm G}/{\rm H} &=& \SO{8,4+n}/ ( \SO{8}\times \SO{4+n} ) \;.
\la{GH}
\ea
The next step is the identification of the gauge group within~${\rm
G}$. According to~\cite{NicSam03a}, an $\SO4$ YM gauging is equivalent
on shell to a CS gauging with gauge group
\ba
\SO4_{\rm gauge} \ltimes {\rm T} \;,
\la{gt}
\ea
where ${\rm T}\equiv{\rm T}_6$ denotes an abelian group of six
translations that transform in the adjoint representation of
$\SO4_{\rm gauge}$. In addition, the precise embedding of this group
into ${\rm G}$ is constrained by the group-theoretical algebraic
constraints on its embedding tensor~\cite{NicSam01b,dWHeSa03},
see~\Ref{susy} below.

To start with, we embed the $\SO{4}$ subgroup, which will be
identified with the YM gauge group, into the compact subgroup ${\rm
H}\subset{\rm G}$ according to
\ba 
\SO4_{\rm gauge}
&\subset& \underbrace{\SO{4}_+ \times  \SO{4}_- \vphantom{X^j_j} }
\times \underbrace{\SO{4}_2 \times \SO{n} \vphantom{X^j_j}  }
\non[1ex] &\subset& \hspace*{2.5em} \SO{8} \hspace*{2.1em}
\times\hspace*{1.5em} \SO{4+n} \qquad\qquad\;, 
\la{G0} 
\ea 
where 
\ba
\SO{4}_{{\rm gauge}} \equiv {\rm diag}\, \Big( \SO{4}_+ \times
\SO{4}_2 \Big) \;,
\la{so4g} 
\ea 
denotes the diagonal subgroup of $\SO{4}_+$ and $\SO{4}_2$. The
$32+8n$ scalars that parametrize the coset space~\Ref{GH} transform as
a bivector $(8_v,4+n)$ under ${\rm H}$. Under $\SO4_{\rm gauge}$ they
decompose into
\ba &&
\left( [ \ft12,\ft12] + 4\cdot [0,0]  \right) \times \left( [
\ft12,\ft12] + n\cdot [0,0]  \right) \non[1ex] &&\qquad = ~ n
\cdot [\ft12,\ft12] + 4n \cdot [0,0] + [0,0] + 4 \cdot
[\ft12,\ft12] + [0,1] + [1,0] + [1,1] \;. \la{spec1} 
\ea
We observe that these are precisely the bosonic representations
appearing in table~\ref{spectrum}. Recalling that the fermions
transform as $(8_c,4+n)$ under ${\rm H}$ it is straightforward to
verify that the fermionic spectrum also comes out correctly:
\ba
&&
\left(
2\cdot [ \ft12,0] + 2\cdot [0,\ft12]  \right) \times
\left(
[\ft12,\ft12] + n\cdot [0,0]  \right)
\non[1ex]
&&\qquad = ~
2 \cdot [1,\ft12] + 2 \cdot [0,\ft12] + 2 \cdot [\ft12 ,1] +
2 \cdot [\ft12 ,0] + 2n \cdot [\ft12 ,0] + 2n \cdot [0,\ft12] \;.
\ea
Moreover, the factors $\SO{4}_-$ and $\SO{n}$ in \Ref{G0} commute with
$\SO4_{\rm gauge}$ and thus represent global symmetries of the gauged
theory. Further identifying
\ba
{\rm SO}(4)_{\rm glob} &\equiv& \SO{4}_- \;,\qquad
{\rm SO}(n)_{\rm glob} ~\equiv~ \SO{n} \;,
\la{global}
\ea
the decomposition according to \Ref{G0} precisely reproduces the
spectrum of representations of table~\ref{spectrum}. Note however,
that the vector degrees of freedom of table~\ref{spectrum} appear
among the scalars in~\Ref{spec1}. These are the Goldstone bosons which
give mass to the associated CS gauge vectors. Accordingly, the gauge
group~\Ref{so4g} is enlarged to~\Ref{gt} by the essentially unique set
of six nilpotent abelian translations ${\rm T}\subset{\rm G}$
transforming in the adjoint representation of $\SO4_{\rm gauge}$. This
part of the gauge group is broken at the groundstate in order to
account for the massive vectors. We have thus identified the group
\Ref{gt} within~${\rm G}$.

The Lagrangian and supersymmetry variations of the most general $D=3$, 
$N=8$ gauged supergravity have been given in~\cite{NicSam01b}. As shown 
there, the theory is completely specified by the coset space~\Ref{GH} 
and the symmetric embedding tensor $\GTh_{\cM\cN}$, which encodes the
minimal coupling of vector fields to scalars according to
\bea D_\mu \CS &\equiv& \left( \dd_\mu +
\GTh_{\cM\cN}\,B^\cM_\mu\,t^\cN \right) \CS \;. \la{DS} \eea 
The matrix $\CS \in {\rm G}=\SO{8,4+n}$ here contains the scalar
fields of the theory; by $t^\cM$ we denote the generators of
${\mathfrak g}={\rm Lie}\,{\rm G}$ acting by left multiplication, with
the curly indices $\CM$, $\CN$ referring to the adjoint representation
of ${\mathfrak g}$. The number of vector fields involved in \Ref{DS}
is equal to the rank of $\GTh_{\cM\cN}$.  Because the embedding tensor
characterizes the theory completely, the task can be reduced to the
identification of the tensor $\GTh_{\cM\cN}$ that correctly reproduces
the gauge group~\Ref{gt} shown above, and at the same time is
compatible with the algebraic constraints \Ref{susy} below, imposed by
supersymmetry. It turns out that there is a unique $\GTh_{\cM\cN}$
that fits all the requirements.

We denote by indices $I, J, \dots$ and indices $r, s, \dots$ the
vector representations of $\SO8$ and $\SO{4+n}$, respectively. The
generators $\{t^\cM\}$ of ${\mathfrak g}$ split into the compact
generators $\{X^{[IJ]}, X^{[rs]}\}$, and the noncompact generators
$\{Y^{Ir} \}$ with commutation relations\footnote{Upon redefining
$X^{rs} \rightarrow - X^{rs}$, the algebra~\Ref{SO} may be written
in the more familiar form $[X^{\CI\CJ},X^{\CK\CL}] =
2\,(\eta^{\CI[\CK}\,X^{\CL]\CJ}-\eta^{\CJ[\CK}\,X^{\CL]\CI})$ with
$\SO{8,4+n}$ vector indices $\CI=(I,r)$, and the metric
$\eta^{\CI\CJ}\equiv (\Gd^{IJ},-\Gd^{rs})$.  The supersymmetry
constraints \Ref{susy} are then equivalent to the total
antisymmetry of the embedding tensor $\GTh_{\CI\CJ,\CK\CL}$ in the
$\SO{8,4+n}$ indices $[\CI\CJ\CK\CL]$.}
\ba
{}[X^{IJ},X^{KL}] &=&
2\,(\delta^{I[K}\,X^{L]J}-\delta^{J[K}\,X^{L]I})\;,\qquad
[X^{IJ},Y^{Kr}] ~=~ -2 \delta^{K[I}\,Y^{J]r}\;,
\non
{}[X^{rs},X^{uv}] &=&
2\,(\delta^{r[u}\,X^{v]s}-\delta^{s[u}\,X^{v]r}) \;,\qquad
[X^{rs},Y^{Ku}] ~=~ -2 \delta^{u[r}\,Y^{Ks]}\;,
\non
{}[Y^{Ir},Y^{Js}] &=& \delta^{IJ}\,X^{rs} + \delta^{rs}\,X^{IJ} \;.
\la{SO}
\ea
In particular, the current~\Ref{DS} decomposes into
\ba
\CS^{-1} D_\mu \CS &\equiv& \ft12 \CQ_\mu^{IJ}X^{IJ} + \ft12
\CQ_\mu^{rs}X^{rs} +  \CP_\mu^{Ir}Y^{Ir} \;.
\la{QP}
\ea
In this basis, the algebraic constraints imposed by supersymmetry on
the embedding tensor $\Theta_{\cM\cN}$ from \Ref{DS} read
\ba
\GTh_{IJ,KL} &=& \GTh_{[IJ,KL]} \;,
\qquad
\GTh_{IJ,rs}~=~\GTh_{Ir,Js}\;,
\qquad
\GTh_{rs,uv}~=~ \GTh_{[rs,uv]} \;,
\non
\GTh_{IJ,Kr} &=& \GTh_{[IJ,K]r} \;,
\qquad
\GTh_{Kr,su} ~=~ \GTh_{K[r,su]}
\;.
\la{susy}
\ea
Let us mention that supersymmetry actually implies a slightly weaker
set of constraints, e.g.\ it also allows for a trace part in $\GTh$,
see~\cite{NicSam01b,dWHeSa03} for details. For our purpose, however,
the constraints \Ref{susy} are sufficient to determine the consistent
embedding tensor.

In order to describe the embedding according to~\Ref{G0}, we further
need to split these indices into $I=(i,\Ti)$ and $r=(\Hi,\Tr)$ with
\ba I, J, \dots &:&\qquad i, j, \dots \in \{1, 2, 3, 4 \} \;,
\quad \Ti, \Tj, \dots \in \{5, 6, 7, 8\} \;, \non r, s, \dots
&:&\qquad \Hi, \Hj, \dots \in \{1, 2, 3, 4 \} \;, \quad \Tr, \Ts,
\dots \in \{1, 2, \dots, n\} \;. \la{indices} \ea 
The indices $\Ti, \Tj, \dots $ are the same as in \Ref{FreundRubin},
and the indices $\Tr,\Ts,\dots$ the same as in table~II. The
generators $\{t^\cM\}$ accordingly decompose into
\ba \mathfrak{g}
&=& \left\{ X^{[ij]}, X^{i\Tj}, X^{[\Ti\Tj]}, X^{[\Hi\Hj]},
X^{\Hi\Tr}, X^{[\Tr\Ts]} \right\} \oplus \left\{ Y^{i\Hj},
Y^{i\Tr}, Y^{\Ti\Hj}, Y^{\Ti\Tr} \right\} \;. \la{so8n} \ea
{}From these we may explicitly build the generators
of~$\mathfrak{so}(4)_{\rm gauge}$ and
the six abelian nilpotent translations~$\mathfrak{t}$ as
\ba
\mathfrak{so}(4)_{\rm gauge} &\equiv&
\left\{ \CJ^{[ij]} \equiv X^{[ij]} + X^{[\Hi\Hj]} \right\}
\;, \non
\mathfrak{t} &\equiv& \left\{ \GT^{[ij]} \equiv
X^{[ij]} - X^{[\Hi\Hj]} + Y^{i\Hj} - Y^{j\Hi}  \right\} \;.
\la{jt}
\ea
It is straightforward to verify that the $\CJ^{[ij]}$ close into
the $\SO{4}$ algebra~\Ref{so4g} while the mutually commuting
generators $\GT^{[ij]}$ transform in the adjoint representation
under $\CJ^{[ij]}$. This is the Lie algebra underlying \Ref{gt}.

Similarly defining vector fields
\ba
C^{[ij]} &\equiv& B^{[ij]} + B^{[\Hi\Hj]}
\;, \qquad
A^{[ij]} ~\equiv~
B^{[ij]} - B^{[\Hi\Hj]} + B^{i\Hj} - B^{j\Hi} \;,
\ea
we start from the following ansatz for the embedding
tensor $\GTh_{\cM\cN}$~\cite{NicSam03a}
\ba
\GTh_{\cM\cN}\, B^\cM_\mu t^{\cN} &=& \ft12 g_1
\left(
C_\mu^{+[ij]}\CT^{+[ij]}-C_\mu^{-[ij]}\CT^{-[ij]}
+ A_\mu^{+[ij]}\CJ^{+[ij]}-A_\mu^{-[ij]}\CJ^{-[ij]} \right)
\non[.5ex]
&& + \ft12 g_2
\left( A_\mu^{+[ij]}\CT^{+[ij]}-A_\mu^{-[ij]}\CT^{-[ij]} \right)
\;,
\la{thetaYM}
\ea
with real constants $g_1$, $g_2$, and where $A^{\pm [ij]}$ denote the
selfdual and anti-selfdual part of $A^{[ij]}$, respectively,
etc. Translating~\Ref{thetaYM} back into the basis~\Ref{SO},
\Ref{so8n}, this embedding tensor takes the form
\ba
\GTh_{ij,kl} &=& (g_2+2g_1) \,\Ge_{ijkl}
\;,\quad
\GTh_{ij,\Hk\Hl} ~=~ -g_2 \,\Ge_{ijkl}
\;,\quad
\GTh_{ij,k\Hl} ~=~ (g_1+g_2) \,\Ge_{ijkl}
\;,\non
\GTh_{\Hi\Hj,\Hk\Hl} &=& (g_2-2g_1) \,\Ge_{ijkl}
\;,\quad
\GTh_{i\Hk,j\Hl} ~=~ -g_2 \,\Ge_{ijkl}
\;,\quad
\GTh_{\Hi\Hj,k\Hl} ~=~ (g_1-g_2) \,\Ge_{ijkl} \;.
\la{thetaex}
\ea
The choice of a relative minus sign between selfdual and anti-selfdual
components in~\Ref{thetaYM}, or equivalently the relative coupling
constant $(-1)$ between the two $\SO{3}$ factors in $\SO{4}$ is
necessary to ensure that terms proportional to $\Gd^{kl}_{ij}$ drop
out in \Ref{thetaex}, such that the supersymmetry
constraints~\Ref{susy} are satisfied for any choice of $g_1$ and
$g_2$. That is, at this stage we still have a class of physically
distinct theories for different choices of $g_1$, $g_2$. We further
emphasize that these constraints harmonize beautifully with the
particular non-semisimple type of gauge group~\Ref{gt}. Indeed,
coupling the diagonal $\SO4$ of \Ref{so4g} requires a nonvanishing
contribution in $\GTh_{IJ,rs}$. By means of \Ref{susy} this induces a
nonvanishing $\GTh_{Ir,Js}$ which in turn precisely corresponds to
coupling the nilpotent contributions of \Ref{jt}. In other words, the
diagonal $\SO4_{\rm gauge}$ from \Ref{so4g} alone is not a consistent
CS gauge group; supersymmetry requires its non-semisimple extension
to~\Ref{gt}.\footnote{However, a consistent gauging with gauge group
$\SO4$ is still possible with a {\it different} embedding,
corresponding to only one of the factors in \Ref{so4g}, cf. eq.~(19)
of \cite{NicSam01b}.}

We may now state the complete bosonic Lagrangian of the
three-dimensional theory given as a gravity coupled CS gauged 
${\rm G}/{\rm H}$ coset space $\sigma$-model
\ba
e^{-1}\CL &=& -\ft1{4} R
+ \ft1{4} g^{\mu\nu}\,\CP^{Ir}_\mu\CP^{Ir}_\nu
- e^{-1}\CL_{\rm CS} - W \;.
\label{LCS}
\ea
The kinetic scalar term is obtained from putting together \Ref{DS},
\Ref{QP}, and \Ref{thetaex}, while the CS term is
\ba
\CL_{\rm CS}&=&
\ft14\Geps^{\mu\nu\rho} B^\cM_\mu \,\GTh_{\cM\cN}\,
\left( \dd_\nu B^\cN_\rho
+ \ft13\,
f^{\cN\cP}{}_{\cL}\,\GTh_{\cP\cK}\,B^\cK_\nu B^\cL_\rho\right)
\;,
\ea
with $\GTh_{\cM\cN}$ from \Ref{thetaex} and the $\SO{8,4+n}$
structure constants from \Ref{SO}. The potential $W$ is given as a
function of the scalar fields as
\ba
W &=&  - \ft1{48} \left(
T_{[IJ,KL]}T_{[IJ,KL]}+ \ft1{4!}\,\Ge^{IJKLMNPQ}\, T_{IJ,KL}T_{MN,PQ}
- 2\, T_{IJ,Kr}T_{IJ,Kr}
\right) \;,
\la{W}
\ea
in terms of the so-called $T$-tensor
\ba T_{IJ,KL} &=&
\CV^{\cM}{}_{\!IJ}\CV^{\cN}{}_{\!KL}\,\GTh_{\cM\cN} \;, \qquad
T_{IJ,Kr} ~=~ \CV^{\cM}{}_{\!IJ}\CV^{\cN}{}_{\!Kr}\,\GTh_{\cM\cN}
\;, \la{T} \ea
where $\CV$ defines the group matrix $\CS$ in the adjoint
representation:
\ba \CS^{-1} t^\cM \CS &\equiv&  \ft12\,
\CV^{\cM}{}_{\!IJ} \,X^{IJ}+ \ft12\, \CV^{\cM}{}_{\!rs}
\,t^{rs}+\CV^{\cM}{}_{\!Ir} \,Y^{Ir} \;. \la{V} \ea 
Hence, like all other terms in \Ref{LCS}, the scalar potential~$W$
depends crucially on the precise form of the embedding
tensor~\Ref{thetaex}. For the fermionic contributions and full
supersymmetry transformations we refer to \cite{NicSam01b}. Here, we
just quote the variations of the gravitinos $\psi^A_\mu$ and fermion
fields $\chi^{\dA r}$ (neglecting cubic spinor terms)
\ba 
\Gd\psi^A_\mu &=& D_\mu \Ge^A - \ft{\I}{48}\GG^{IJKL}_{AB}
T_{IJ,KL}\Gg_\mu\Ge^B \;,
\nonumber\\
\Gd\chi^{\dA r} &=& \left(\ft{\I}2\GG^I_{\!\!A\dA}\,\,/\!\!\!\!\CP^{Ir}
-\ft1{12}\GG^{IJK}_{A\dA} T_{IJ,Kr}\right) \Ge^A
\;,
\ea
with $\SO8$ $\GG$-matrices $\GG^I_{\!\!A\dA}$. These variations are
likewise expressed in terms of the $T$-tensor from \Ref{T} and may
serve as BPS equations for bosonic solutions. In particular, they show
that an AdS ground state preserving all supersymmetries requires
$T_{IJ,Kr}=0$. Recall that we seek that theory whose groundstate at
the origin $\CS= {\mathbb I}$ precisely corresponds to the six-dimensional
AdS$_3\times S^3$ background with full $N=(4,4)$ supersymmetry. Since
$T$ at this point reduces to the embedding tensor $\GTh$, together
this imposes a nontrivial relation between the constants $g_1$, $g_2$
in \Ref{thetaex}
\ba
\GTh_{IJ,Kr}&=& 0 \qquad \Longrightarrow \qquad
\GTh_{ij,k\Hl}~=~ 0 \qquad \Longrightarrow \qquad g_2=-g_1 \;.
\la{g1g2}
\ea
That is, existence of a maximally supersymmetric AdS groundstate
eventually fixes the ratio $g_1/g_2$, such that the final theory is
completely determined up to an overall coupling constant which may be
expressed in terms of the AdS length $L_0$ at the origin as
$g_1=-g_2=1/L_0$. At this point, the gauge group~\Ref{gt} breaks down
to its compact part $\SO4_{\rm gauge}$; i.e.\ the background isometry
group which organizes the spectrum of fluctuations around this point
is the desired $\SU{2|1,1}_L\times\SU{2|1,1}_R$. The vector fields
corresponding to the translational part of \Ref{gt} acquire a mass in
accordance with table~\ref{spectrum}.

We have thus succeeded in constructing a three-dimensional $N=8$
supersymmetric theory with an $N=(4,4)$ supersymmetric AdS$_3$ ground
state at the origin of the scalar potential which reproduces the correct
symmetries and the field content of table~\ref{spectrum}. As a further
check, one may explicitly compute quadratic fluctuations of the
potential~\Ref{W} around the ground state $\CS={\mathbb I}$ to obtain 
the scalar masses. Indeed, after some calculation, one confirms the 
scalar mass spectrum obtained from table~\ref{spectrum} via the 
standard relation
\ba
m^2L_0^2 &=& \Delta(\Delta-2) \;,
\ea
together with vanishing masses for the Goldstone bosons. Furthermore,
half of the 12 CS vector fields acquire mass in a three-dimensional
variant of the Brout-Englert-Higgs effect: linearizing their first
order field equations around the origin of the scalar potential leads
to
\ba
\Ge_{\mu\nu\rho} \CA^{\pm [ij]\,\nu\rho} &=&
\mp 4 g_1\,
\left(A^{\pm[ij]}_\mu - C^{\pm[ij]}_\mu\right)
\;,\qquad \CC^{\pm[ij]}_{\mu\nu}  ~=~ 0 \;.
\la{VClin1}
\ea
After the obvious redefinitions, these equations are identical
with \Ref{vclin1}. Alternatively, we recall from~\cite{FiNiSa02}
that the vector mass matrix around the origin may be directly
obtained from the projection of the embedding tensor
$\GTh_{Ir,Js}$ onto its noncompact directions, i.e.\ here from the
components $\GTh_{i\Hk,j\Hl}$ in~\Ref{thetaex}. Via the
mass-dimension relation $\Delta=1+|mL_0|$ for three-dimensional
vectors, we find precise agreement with the masses of
table~\ref{spectrum}.  We may finally employ the results
of~\cite{NicSam03a} to obtain the on-shell equivalent version of
the theory \Ref{LCS} which describes an $\SO4$ YM gauging. To this
end, we split off the scalars $\phi_{ij}$ corresponding to the six
nilpotent directions of ${\rm T}_6$ from the group matrix $\CS$
according to
\ba
\CS &\equiv& e^{\phi_{ij}\CT^{[ij]}}\,\TCS \;,
\ea
where $\TCS$ now combines the remaining $26+8n$ scalar
fields. Correspondingly, we define its currents
\ba
\TCQ_\mu + \TCP_\mu &\equiv&
\TCS^{-1}\left(\dd_\mu ~+~ \ft12 g_1
(A_\mu^{+[ij]}\CJ^{+[ij]}-A_\mu^{-[ij]}\CJ^{-[ij]}) \right)
\TCS
\;,
\ea
and the matrix $\TCV^{[\underline{ij}]}{}_\cM$ as
\ba
\TCS^{-1} \CT^{[ij]} \TCS &\equiv&
\ft12\, \TCV^{[\underline{ij}]}{}_{\!IJ} \,X^{IJ}+
\ft12\, \TCV^{[\underline{ij}]}{}_{\!rs} \,X^{rs}+
\TCV^{[\underline{ij}]}{}_{\!Ir} \,Y^{Ir}
\;.
\ea
Eliminating the scalar fields $\phi_{ij}$ together with the CS vector
fields $C_\mu^{[ij]}$ from the above Lagrangian~\Ref{LCS} as described
in~\cite{NicSam03a}, turns this theory into a YM type gauged supergravity 
with $\SO4$ YM vector fields~$A_\mu^{[ij]}$ and the Lagrangian
\ba
e^{-1}\TCL &=& -\ft1{4} R
+ \ft1{4} g^{\mu\nu}\,G_{Ir,Js}\,\TCP^{Ir}_\mu\TCP^{Js}_\nu
+\ft1{16} \,M_{[ij],[kl]}\,\CA^{[ij]\mu\nu}\CA^{[kl]}_{\mu\nu}
- e^{-1}g_1\! \left(\TCL^{(+)}_{\rm CS}-\TCL^{(-)}_{\rm CS}\right)
\non[1ex]
&&{}
+\ft12 e^{-1} \,
\Geps^{\mu\nu\rho}\,M_{[ij],[kl]}\,
\TCV^{[\underline{kl}]}{}_{Ir}\,\CA^{[ij]}_{\mu\nu}\,\TCP^{Ir}_\rho
- W \;,
\label{LYM}
\ea
with
\ba
G_{Ir,Js} &\equiv& \delta_{IJ}\delta_{rs} -
\TCV^{[\underline{ij}]}{}_{Ir} M_{[ij],[kl]}
\TCV^{[\underline{kl}]}{}_{Js}
\;,\qquad
M_{[ij],[kl]} ~\equiv~
(\TCV^{[\underline{ij}]}{}_{Ir}\TCV^{[\underline{kl}]}{}_{Ir} )^{-1}
\;,\non[1ex]
\TCL^{(\pm)}_{\rm CS}&=&
\ft18\, \Geps^{\mu\nu\rho} A^{\pm[ij]}_\mu\,
\left( \dd_\nu A^{\pm[ij]}_\rho
\pm \ft13 g_1 A^{\pm[ik]}_\nu A^{\pm[jk]}_\rho\right)
\;.
\nn
\ea
Notably, the scalar potential $W$ of \Ref{LYM} coincides with the one
derived above~\Ref{W}. In particular, the scalar mass spectrum around
the origin $\TCS={\mathbb I}$ coincides with the one of \Ref{LCS} and thus 
with table~\ref{spectrum}. On the other hand, the vector field equations
obtained from linearizing \Ref{LYM} around $\TCS={\mathbb I}$ give
\ba
\nabla^\nu \CA^{\pm [ij]}_{\mu\nu} &=&
\pm g_1 \Ge_{\mu\nu\rho}\,\CA^{\pm[ij] \nu\rho}
\;,
\ea
and thus precisely reproduce~\Ref{vclin2}. Summarizing, with
\Ref{LCS} and \Ref{LYM} we have constructed the two equivalent
versions of the three-dimensional theory that yield the full nonlinear
extension of the linearized field equations \Ref{vclin1} and
\Ref{vclin2}, respectively.

\mathon
\section{Coupling the KK towers}
\mathoff

In the previous section we have constructed a theory coupling the
two supermultiplets 
\ba 1\cdot ({\bf 3},{\bf 3})_S  + n \cdot
({\bf 2},{\bf 2})_S \;,
\ea 
to the massless supergravity multiplet, and hence succeeded in
combining the lowest levels of the three KK towers of six-dimensional
$N=(2,0)$ supergravity on AdS$_3\times S^3$. Apart from the spin-2
tower, the entire KK spectrum is given by two infinite towers of
spin-$1$ multiplets
\ba \CH_{{\rm
spin-}1} &=& \sum_{k\ge 2} ({\bf k+1},{\bf k+1})_S ~+~ n \cdot
\sum_{l\ge1} ({\bf l+1},{\bf l+1})_S \;, \la{KKspectrum} \ea
transforming in the singlet and the vector representation of $\SO{n}$,
respectively. In this section, we will show that extending the above
construction we may in fact construct a three-dimensional theory that
comprises the entire spectrum~\Ref{KKspectrum} of spin-$1$
multiplets. It is obtained by gauging the $N=8$ theory with the coset
space given in \Ref{GI}. If one wishes, one can ``regularize'' the
infinite component theory by considering only a finite but arbitrarily
large number of multiplets.

Counting the number of bosonic degrees of freedom in \Ref{KKspectrum}
via table~\ref{spin1}, indeed reproduces the infinite dimensional
coset \Ref{GI}, i.e.\ gives agreement for each value of $k$ and $l$ 
separately. It remains to determine the CS gauge group and 
its embedding. According to~\cite{NicSam03a}, this group, which 
we denote by $\SO4_{\rm gauge} \ltimes(\hat{{\rm T}}_\infty,{\rm T}_6)$, 
is obtained from exponentiating an extension of the
non-semisimple algebra~\Ref{jt} described above, by a set of
additional nilpotent generators $\hat{{\mathfrak{t}}}_\infty$
corresponding to the massive vector fields appearing
in~\Ref{KKspectrum}. More precisely, the generators of
$\hat{{\mathfrak{t}}}_\infty$ should transform like the massive vector
fields in the
\ba
\sum_{k\ge3} \left( \left[\ft{k-2}2,\ft{k}2\right]+
\left[\ft{k}2,\ft{k-2}2\right]\right)  ~+~
\sum_{l\ge2} \left(
n\cdot\left[\ft{l-2}2,\ft{l}2\right]+
n\cdot\left[\ft{l}2,\ft{l-2}2\right]\right)
\;,
\label{massiveV}
\ea
under $\SO4_{\rm gauge}$ (cf.\ table~\ref{spin1}) and close into
${\mathfrak{t}}$. To illustrate the embedding of the gauge group and
the global symmetries in the group~${\rm G}$, let us first consider
the decomposition of its maximal compact subgroup~${\rm H}$ into

\vspace*{-1.2em}
{\scriptsize
\ba
\underbrace{\SO{4}_+ \times  \SO{4}_- \vphantom{X^j_j} }
\times \underbrace{\SO{4}_2\vphantom{X^j_j}  }
\times \dots \times
\underbrace{\SO{4}_{k}\vphantom{X^j_j}  }\;
\times \dots \;\;\;
\times \underbrace{\SO{n}_1\vphantom{X^j_j}  } \times
\underbrace{\SO{n}_2 \times \SO4'_2 \vphantom{X^j_j}  }\;\times \dots
\times \underbrace{\SO{n}_l \times \SO4'_l
\vphantom{X^j_j}  }\;\times \dots
\non[.4ex]
\hspace*{2.9em} \SO{8}   \hspace*{2.4em}
\times\; \underbrace{ \SO{4}  \times
\dots \times \hspace*{.4em} \SO{k^2}
\; \times \dots \vphantom{X^j_j} }
\times\; \underbrace{ \SO{n}  \hspace*{1.4em} \times
\hspace*{1.4em} \SO{4n}  \hspace*{1.4em}
\times \;\;\;\dots \;\;\;\times \hspace*{1.4em}
\SO{n l^2} \;\times \;\; \dots  \vphantom{X^j_j} \;\;}
\non[.4ex]
\SO8 \hspace*{2.4em}\times \hspace*{3.3em}
{\rm SO}({\textstyle \sum_{k\ge 2} k^2})
\hspace*{4em}\times\hspace*{8.5em}
{\rm SO}({\textstyle n \sum_{l\ge1} l^2})
\hspace*{9em}
\nn
\ea
}
\vspace*{-4.5em}
\be
\la{G08}
\ee
with the specific embeddings
\ba
\SO4_k ~\subset~ \SO{k^2} \;&:&\quad
k^2 \rightarrow \left[\ft{k\mis1}2,\ft{k\mis1}2\right]
\;,
\non[.5ex]
\SO{n}_l \times \SO4'_l ~\subset~ \SO{n l^2} &:&\quad
nl^2 \rightarrow \left[n,\ft{l\mis1}2,\ft{l\mis1}2\right] \;.
\ea
We then define the group $\SO{4}_{\rm gauge}$ and
the global symmetry groups as the diagonal subgroups
\ba
\SO{4}_{\rm gauge} &\equiv&
\diag\!\left(\SO4_+\times
\SO{4}_2 \times \SO{4}_3 \times \dots \times
\SO{4}'_2 \times \SO{4}'_3 \times \dots \; \right) \;,
\non[.5ex]
\SO{4}_{\rm glob} &\equiv& \SO4_- \;, \qquad
\SO{n}_{\rm glob} ~\equiv~
\diag\!\left(\SO{n}_1\vphantom{\SO{4}'_3}
\times \SO{n}_2 \times \dots  \; \right) \;.
\la{gloloc}
\ea
Working out the products of the relevant representations, viz.
\ba
&& \left( 1\cdot \left[\ft12 ,\ft12 \right] + 4\cdot [0,0] \right)
\times
   \left[\ft{k-1}2, \ft{k-1}2 \right] \qquad ({\rm bosons})
\;,
\non
&& \left( 2\cdot\left[\ft12 ,0 \right] + 2\cdot [0,\ft12] \right)
\times
   \left[\ft{k-1}2, \ft{k-1}2 \right] \qquad ({\rm fermions}) \nonumber
\;,
\ea
it is straightforward to verify that under these subgroups the fields
reproduce the correct representation content as given by
\Ref{KKspectrum} together with table~\ref{spin1}. Again, the vector
degrees of freedom show up through the associated Goldstone scalars,
transforming in the same representations as the massive vector
fields~\Ref{massiveV}.

For transparency, we will now describe in detail the theory which
couples the distinguished lowest spin-1 multiplet $({\bf 3,3})_S$ to a
single additional higher spin-$1$ multiplet, say, $({\bf k\pls1},{\bf
k\pls1})_S$ from the $\SO{n}$ singlet tower.  The extension to an
arbitrary number of these multiplets and multiplets from the $\SO{n}$
vector tower is straightforward.

Let us therefore consider the coset space
\ba
{\rm G}/{\rm H} &=& \SO{8\,,\,4+k^2}/
\left( \SO{8}\times \SO{4+k^2} \right) \;,
\la{GHk}
\ea
with indices split according to $I=(i,\Ti)$ and $r=(\Hi,ab)$ with $a,
b, \dots = 1, \dots, k$, extending \Ref{indices}. The generators
$\{t^\cM\}$ of $\SO{8,4+k^2}$ accordingly decompose as
\ba
\mathfrak{g} &=&
\left\{ X^{[ij]}, X^{i\Tj}, X^{[\Ti\Tj]},
X^{[\Hi\Hj]}, X^{\Hi,ab}, X^{ab,cd}
\right\} \oplus
\left\{ Y^{i\Hj}, Y^{i,ab}, Y^{\Ti\Hj}, Y^{\Ti,ab} \right\} \;.
\la{so84k}
\ea
The commutation relations can be read off from \Ref{SO}.
For the $\SO{k^2}$ subgroup, we thus have
\be
[ X^{ab,cd} , X^{ef,gh} ]
   = \eta^{cd,ef} X^{ab,gh} - \eta^{cd,gh} X^{ab,ef} -
   \eta^{ab,ef} X^{cd,gh} + \eta^{ab,gh} X^{cd,ef} \;,
\ee
where the metric $\eta_{ab,cd} \equiv \eta_{ac}\eta_{bd}$ serves to
lower and raise indices, such that
\be
\eta_{ab,ef} \eta^{ef,cd} = \Gd_a^c \Gd_b^d \;,
\ee
and the tensor $\eta_{ab}$ is the quadratic invariant of the $k$-dimensional
representation of $\SO3$; it is symmetric for bosonic representations,
i.e.\ odd $k$, and skew-symmetric for fermionic representations, i.e.\
even $k$.

In line with our general arguments above we now seek a theory with
CS gauge group
\ba
\SO{4}_{\rm gauge} \ltimes
\left(\hat{{\rm T}}^{(k)}, {\rm T}_6\right) \subset \SO{8, 4+k^2}
\;,
\la{gtt}
\ea
with $\SO{4}_{\rm gauge}$ from \Ref{gloloc}, extending \Ref{gt} by
$2(k^2\mis1)$ generators transforming in the
$\left[\ft{k-2}2,\ft{k}2\right]+\left[\ft{k}2,\ft{k-2}2\right]$ under
$\SO{4}_{\rm gauge}$ and closing into~${\rm T}_6$, in order to
correctly describe a theory with $\SO4$ YM gauging and $2(k^2\mis1)$
massive vector fields.
The generators of~$\mathfrak{so}(4)_{\rm gauge}$ are defined as the
extension of the previous $\SO4$ to the new diagonal $\SO4$ subgroup
embedded into $\SO{8,4+k^2}$ according to \Ref{gloloc}, viz.
\be
\mathfrak{so}(4)_{\rm gauge} =
\left\{ \CJ^{[ij]} \equiv X^{[ij]} + X^{[\Hi\Hj]}
+ \ft12\, \Gz^{+(k)}_{ij\;ac}\eta_{bd}\, X^{ab,cd}
+ \ft12\, \Gz^{-(k)}_{ij\;bd}\eta_{ac}\, X^{ab,cd}
 \right\} \;.
\la{jk}
\ee
By $\Gz^{\pm(k)}_{ij}$, we here denote
the generators of $\SO4$ in the spin $\left[\ft{k-1}2,0\right]$
and spin $\left[0,\ft{k-1}2\right]$ representation, respectively.
Accordingly, $\Gz^{\pm(k)}_{ij\;ab}$ is symmetric in $ab$ for fermionic
(even $k$) and skew-symmetric for bosonic (odd $k$) representations.
Explicit expressions for these generators can be constructed in terms
of Clebsch-Gordan coefficients. The $\SO{3}^\pm$ commutation relations
in this representation are
\ba\label{SO3z}
\left[\, \Gz^{\pm(k)}_{ij} , \Gz^{\pm(k)}_{\vl mn}\, \right] ~=~
2\left( \delta^\vl_{i[m} \Gz^{\pm(k)}_{n]j} -
\delta^\vl_{j[m} \Gz^{\pm(k)}_{n]i} \right) \;,
\quad &&
\Gz^{\pm(k)}_{ij} ~=~ \pm \ft12\,\Ge^{ijmn}\,\Gz^{\pm(k)}_{mn}\;,
\non[1ex]
\tr\! \left(\Gz^{(\pm k)}_{ij}\,\Gz^{(\pm k)}_{ji} \right) ~=~
k\,(k^2-1)
\;,
\quad &&
\ea
in obvious matrix notation. Let us also record the relation
\ba\label{zz}
\Gz^{\pm(k)}_{im \, ac}\, \Gz^{\pm(k)}_{mj\, cb} =
\ft14 (k^2 -1) \Gd_{ij} \eta_{ab} +  \Gz^{\pm(k)}_{ij \, ab}
\;,
\ea
which follows from \Ref{SO3z} and the (anti-)selfduality of the
$\Gz$'s. 

The generators of the translation subgroup $(\hat{T}^{(k)},T_6)$ of
\Ref{gtt} are given by
\be
\mathfrak{t}\equiv \left\{ \GT^{[ij]} \equiv
X^{[ij]} - X^{[\Hi\Hj]} + Y^{i\Hj} - Y^{j\Hi}  \right\} \;,
\ee
which is the same as before, and by
\be
\hat\mathfrak{t}^{(k)} = \left\{
\left({\mathbb
P}_{\left[\frac{k\mis2}2,\frac{k}2\right]}\right)_{i\,ab,j\,cd}
\left( X^{\Hj \, cd} - Y^{j \, cd} \right) \right\} \oplus \left\{
\left({\mathbb
P}_{\left[\frac{k}2,\frac{k\mis2}2\right]}\right)_{i\,ab,j\,cd}
\left( X^{\Hj \, cd} - Y^{j \, cd} \right)
\right\}
\;,
\label{tk}
\ee
where
\ba
\left({\mathbb P}_{\left[\frac{k\mis2}2,\frac{k}2\right]}
\right)_{i\,ab,j\,cd} &=&
\ft{k^2-1}{4k^2}\,\eta_{ac}\eta_{bd} \, \delta_{ij}
+\ft{k+1}{2k^2}\,\zeta^{+(k)}_{ij\;ac}\,\eta_{bd}
-\ft{k-1}{2k^2}\,\eta_{ac}\,\zeta^{-(k)}_{ij\;bd}
-\ft{1}{k^2}\,\zeta^{+(k)}_{im\;ac}\,\zeta^{-(k)}_{mj\;bd} \;,
\non
\left({\mathbb P}_{\left[\frac{k}2,\frac{k\mis2}2\right]}
\right)_{i\,ab,j\,cd} &=&
\ft{k^2-1}{4k^2}\,\eta_{ac}\eta_{bd} \, \delta_{ij}
-\ft{k-1}{2k^2}\,\zeta^{+(k)}_{ij\;ac}\,\eta_{bd}
+\ft{k+1}{2k^2}\,\eta_{ac}\,\zeta^{-(k)}_{ij\;bd}
-\ft{1}{k^2}\,\zeta^{+(k)}_{im\;ac}\,\zeta^{-(k)}_{mj\;bd} \;,
\nn
\ea
are the projectors onto the $\left[\frac{k-2}2,\frac{k}2 \right]$ and
$\left[ \frac{k}2,\frac{k-2}2\right]$ representations, respectively,
in the tensor product
$\left[\frac12,\frac12\right]\times\left[\frac{k-1}2,\frac{k-1}2\right]$
of ${\SO4}_{\rm gauge}$. The projector properties are most easily
verified by writing the above projectors as products of the
corresponding operators for the chiral product $\left[\ft12\right]
\times \left[\ft{k-1}2\right] = \left[\frac{k}2\right] \oplus
\left[\frac{k-2}2\right]$
\ben
\Big({\mathbb P}_{\left[\frac{k}2\right]}\Big)_{ij,ab} ~\equiv~
\ft{k+1}{2k}\, \Gd_{ij}\eta_{ab}
- \ft1{k}\, \zeta^{(k)}_{ij\;ab}
\;\;,\qquad
\left({\mathbb P}_{\left[\frac{k-2}2\right]}\right)_{ij,ab} ~\equiv~
\ft{k-1}{2k} \Gd_{ij}\eta_{ab} + \ft1{k} \zeta^{(k)}_{ij\;ab}
\;\;,
\een
and by use of \Ref{zz}. With the relation
\ben
\left[ X^{\Hi\, ab} - Y^{i\, ab} , X^{\Hj \, cd} - Y^{j\, cd} \right] =
\eta^{ab,cd} \left( X^{ij} - X^{\Hi\Hj} + Y^{i\Hj} - Y^{j\Hi} \right)
\;,
\een
it is easy to show that
\be
[\, \hat\mathfrak{t}^{(k)} , \hat\mathfrak{t}^{(k)}\,]
\subset \mathfrak{t} \;,\qquad
[\, \hat\mathfrak{t}^{(k)} , \hat\mathfrak{t}^{(l)}\,] = 0 \quad
\mbox{for}\;\;k\neq l\;,
\qquad
[\, \hat\mathfrak{t}^{(k)} , \mathfrak{t}\,] = 0
\;.
\ee
Thus, even taking into account an infinite number of translation
subalgebras, the full gauge algebra still has a rather simple
structure.

The embedding tensor \Ref{thetaex} acquires the additional
components 
\ba \GTh^{(k)}_{ij,ab\,cd} &=&
-\GTh^{(k)}_{\Hi\Hj,ab\,cd} ~=~ \GTh^{(k)}_{i\Hj,ab\,cd} ~=~
g_1\,\Gz^{+(k)}_{ij\;ac}\,\eta_{bd} -
g_1\,\Gz^{-(k)}_{ij\;bd}\,\eta_{ac} \;, \non[1ex]
\GTh^{(k)}_{i\,ab,j\,cd} &=& \GTh^{(k)}_{\Hi\,ab,\Hj\,cd} ~=~
-\GTh^{(k)}_{i\,ab,\Hj\,cd} ~=~
g_1\,\Gz^{+(k)}_{ij\;ac}\,\eta_{bd} -
g_1\,\Gz^{-(k)}_{ij\;bd}\,\eta_{ac} \;, 
\la{thetaexk} 
\ea 
which are obviously compatible with the algebraic
constraints~\Ref{susy} imposed by supersymmetry (with antisymmetry
under interchange of the indices $i,j$ and the pairs $ab$ and $cd$,
each pair being regarded as a single ${\rm SO}(k^2)$ index). Moreover,
they do not obstruct the existence of an $N=(4,4)$ supersymmetric AdS
groundstate~\Ref{g1g2} if we keep $g_1 = -g_2$ as we did in
\Ref{thetaex}. The components in the first line of \Ref{thetaexk} can
be read off directly from \Ref{jk} (keeping in mind the relative
factor $(-1)$ between the two $\SO3$ factors in $\SO4$), and give rise
to the generalization of $\CJ^{[ij]}$ from~\Ref{jk}. The
remaining components, i.e.\ the second line in \Ref{thetaexk} lead to
the additional contribution in \Ref{thetaYM}
\ba
\GTh^{(k)}_{\cM\cN} \, B^\cM_\mu t^{\cN} &=&
\GTh^{(k)}_{i\,ab,j\,cd} \left(B^{\Hi\,ab}-B^{i\,ab}\right)
\left(X^{\Hj\,cd}-Y^{j\,cd}\right) \;. 
\la{thetaYMk} 
\ea 
These components can be determined in two {\it a priori} different
ways. On the one hand, they are related to the first line of
\Ref{thetaexk} by supersymmetry \Ref{susy}, i.e.\ complete
antisymmetry in the $\SO{8,4+k^2}$ indices, implying
e.g.~$\GTh^{(k)}_{i\,ab,j\, cd}=\GTh^{(k)}_{ij,ab\, cd}$. On the other
hand, their values are proportional to the difference between the
two projectors from~\Ref{tk}
\ba 
\GTh^{(k)}_{i\,ab,j\,cd} &=& g_1 k\, \Big(
{\mathbb P}_{\left[\frac{k}2,\frac{k\mis2}2\right]} - {\mathbb
P}_{\left[\frac{k\mis2}2,\frac{k}2\right]} \Big)_{i\,ab,j\,cd}
\;\;,
\la{P-P} 
\ea 
again featuring the relative factor of $(-1)$ between the different
``chiralities'' that we saw already in \Ref{thetaYM}. This remarkable
coincidence guarantees that \Ref{thetaYMk} picks out precisely
$2(k^2\mis1)$ vector fields from the a priori $4k^2$ fields
$(B^{\Hi\,ab}-B^{i\,ab})$. Likewise, the combinations
\ba 
\hat{\CT}_{i\,ab} &\equiv&
\GTh^{(k)}_{i\,ab,j\,cd}\,(X^{\Hj\,cd}-Y^{j\,cd}) \;, 
\ea
appearing in~\Ref{thetaYMk} correspond to a projection of the $4k^2$
nilpotent generators $(X^{\Hj\,cd}-Y^{j\,cd})$ onto a subset of
$2(k^2\mis 1)$ generators, which span~\Ref{tk}. Had supersymmetry
\Ref{susy} imposed another value for $\GTh^{(k)}_{i\,ab,j\, cd}$, the
minimal coupling \Ref{thetaYMk} would have involved too many vector
fields and generators. The CS gauge group identified by~\Ref{thetaexk}
precisely realizes the desired algebra~\Ref{gtt}. Again
supersymmetry matches beautifully with the algebraic structure.

The Lagrangian of the theory coupling the higher spin-$1$ multiplet
$({\bf k\pls1},{\bf k\pls1})_S$ is then given by \Ref{LCS}--\Ref{V}
with the coset space ${\rm G}/{\rm H}$ from \Ref{GHk} and the
embedding tensor~$\GTh_{\cM\cN}$ by \Ref{thetaex}, \Ref{thetaexk}. As
a first non-trivial check one may compute the vector mass spectrum
around the origin, encoded in the eigenvalues of
$\GTh^{(k)}_{i\,ab,j\,cd}$, see~\cite{FiNiSa02} and the discussion
after \Ref{VClin1} above. Indeed, from \Ref{P-P} one finds the masses
$mL_0=\pm k$, reproducing the spectrum $\Delta=1+|mL_0|$ of
table~\ref{spin1}. Finally, upon eliminating the scalar fields
corresponding to $\CT^{[ij]}$, $\hat{\CT}_{i\,ab}$
following~\cite{NicSam03a}, one obtains the equivalent formulation of
this theory as an $\SO4$ YM gauge theory with the vector fields
$A^{[ij]}$ promoted to propagating YM vector fields, and the massive
CS vector fields $\GTh^{(k)}_{i\,ab,j\,cd}\,(B^{\Hj\,cd}-B^{j\,cd})$.

The theory describing the entire spectrum~\Ref{KKspectrum} is
straightforwardly constructed starting from the coset~\Ref{GI} and
summing over the additional contributions~\Ref{thetaexk} of the
embedding tensor $\GTh$ for the different $k$. Note that all
$\GTh^{(k)}$ in \Ref{thetaexk} act in different sectors;
consequently, there are no divergent infinite sums of any kind in
the limit of infinitely many multiplets. Similar comments apply to
the multiplets from the spin-$1$ tower in the vector
representation of $\SO{n}$.

\section{Conclusions}

It is rather striking that the complete Lagrangian coupling an arbitrary
number of multiplets from the spin-$1$ KK towers can be cast into the
simple form of \Ref{LCS}
\ba
e^{-1}\CL &=& -\ft1{4} R
+ \ft1{4} g^{\mu\nu}\,\CP^{Ir}_\mu\CP^{Ir}_\nu
- e^{-1}\CL_{\rm CS} - W  + \CL_{\rm ferm}\;,
\label{LCSa}
\ea
with all the complexity encoded in the coset space structure of
\Ref{GI} and the precise form of the embedding
tensor~\Ref{thetaex},~\Ref{thetaexk}.  The complete fermionic
Lagrangian as well as the supersymmetry transformation rules are
obtained from~\cite{NicSam01b} upon using this explicit form of the
embedding tensor.

It remains an open problem whether the KK tower of massive spin-$2$
supermultiplets can be incorporated in the effective three-dimensional
theory in a similar fashion. This would amount to casting the {\em
full} Kaluza-Klein theory into a single $D=3$ supergravity with an
infinite dimensional irreducible coset space and in particular allow
us to address the issue of consistent truncations to finite subsectors
directly within the three-dimensional theory.

\begin{table}[bt]
{\small
\begin{center}
\begin{tabular}{crcccc}\hline
$\Delta$ & $s_0$ & & ${\rm SO}(4)_{\rm gauge} $ &
${\rm SO}(4)_{\rm glob}$ &
\# dof
\\ \hline\hline
$p$ & $1$&  & $\left[\ft{p-1}2,\ft{p+1}2\right]$ & $[0,0]$ & $p(p\pls2)$
\\
$p+\ft12$ &$\ft32$& & $\left[\ft{p-1}2,\ft{p}2\right]$ &
$\left[0,\ft12\right]$  & $2p(p\pls1)$
\\
$p+\ft12$&$\ft12$&  & $\left[\ft{p-2}2,\ft{p+1}2\right]$ &
$\left[\ft12,0\right]$  &  $2(p\mis1)(p\pls2)$
\\
$p+1$ & $1$&&$\left[\ft{p-2}2,\ft{p}2\right]$&
$\left[\ft12,\ft12\right]$  & $4(p^2-1)$
\\
$p+1$ & $2$ && $\left[\ft{p-1}2,\ft{p-1}2\right]$ & $[0,0]$  & $p^2$
\\
$p+1$ & $0$ & & $\left[\ft{p-3}2,\ft{p+1}2\right]$ & $[0,0]$  & $p^2-4$
\\
$p+\ft32$ & $\ft32$&  & $\left[\ft{p-2}2,\ft{p-1}2\right]$ &
$\left[\ft12,0\right]$  &
$2p(p\mis1)$
\\
$p+\ft32$ & $\ft12$ & & $\left[\ft{p-3}2,\ft{p}2\right]$ &
$\left[0,\ft12\right]$  &  $2(p\pls1)(p\mis2)$
\\
$p+2$ &  $1$ & & $\left[\ft{p-3}2,\ft{p-1}2\right]$  & $[0,0]$ & $p(p\mis2)$
\\
\hline
\end{tabular}
\caption{\small Spin-2 multiplet $({\bf p},{\bf p\pls2})_S$ of
$\SU{2|1,1}_L\times\SU{2|1,1}_R$. The conjugate multiplet 
$({\bf p\pls2},{\bf p})_S$ is obtained by $s_0\rightarrow-s_0$, and
$[j_1,j_2]\rightarrow[j_2,j_1]$ under ${\rm SO}(4)_{\rm gauge}$ and
${\rm SO}(4)_{\rm glob}$.  }
\la{spin2}
\end{center}}
\end{table}

We conclude with some intriguing hints that an extension to the full 
KK theory actually exists. The massive spin-2 KK tower contains 
the representations 
\ba
\CH_{{\rm spin-}2} &=& \sum_{p\ge2}\, ({\bf p},{\bf p+2})_S ~+~
\sum_{p\ge2}\, ({\bf p+2},{\bf p})_S \;, \la{KKspectrum2} \ea
with the massive spin-2 multiplet $({\bf p},{\bf p\pls2})_S$ given in
table~\ref{spin2}. The numbers of bosonic and fermionic degrees of
freedom are separately equal $8(p^2-1)$. This suggests that we further
enlarge the coset~\Ref{GI} to a coset space ${\rm G}/{\rm H}$ with the
group
\ba
{\rm G} &=&
{\rm SO}\Big(8\;,\, { \sum_{k\ge 2} k^2} +
n {\sum_{l\ge1} l^2}
+ 2 { \sum_{p\ge2} (p^2-1)}  \Big) \;,
\la{G888}
\ea
and {\rm H} its maximal compact subgroup. With the specific embeddings
\ba
\SO4_p ~\subset~ \SO{2(p^2\mis1)} \;&:&\quad
2\,(p^2-1) \rightarrow
\left[\ft{p\mis2}2,\ft{p}2\right]+
\left[\ft{p}2, \ft{p\mis2}2\right]
\;,
\ea
one may define the group $\SO4_{\rm gauge}$ as the diagonal of
\Ref{gloloc} and the additional $\SO4_p$ groups. It is then
straightforward to verify that the representation content of
\Ref{G888} indeed reproduces~\Ref{KKspectrum2} with table~\ref{spin2}
in the fermionic and the scalar sectors. The construction of a
consistent gauged supergravity with this gauge group, however, is less
obvious than the one presented above. In particular, it remains an
open question if the spin-$2$ fields may be described by a coset space
theory~\Ref{G888} and acquire masses by some as yet undiscovered
version of the Brout-Englert-Higgs mechanism, or if this theory
requires explicit extra couplings (which would vitiate the economy and
beauty of the present scheme to some extent). These issues in turn
hinge on the question whether there exists a novel type of duality
between symmetric tensors and scalars in three dimensions which would
generalize the well-known scalar vector duality. We hope to come back
to these questions in the near future.

\subsection*{Acknowledgements}

We wish to thank M.~Berg and M.~Trigiante for useful discussions, 
and A.~Sagnotti for alerting us to refs.~\cite{GivPor90}.
This work is partly supported by EU contract HPRN-CT-2000-00122 and
HPRN-CT-2000-00131.

\renewcommand{\baselinestretch}{1}
{\small

%\bibliographystyle{h-physrev3}
%\bibliographystyle{Jopt2}
%\bibliography{bib}

\begin{thebibliography}{10}

\bibitem{Mald97}
J.~Maldacena, { The large ${N}$ limit of superconformal field theories and
  supergravity},  { Adv. Theor. Math. Phys.} { 2} (1998) 231--252,
[\href{http://xxx.lanl.gov/abs/hep-th/9711200}{{\tt hep-th/9711200}}].
%%CITATION = HEP-TH 9711200;%%.

\bibitem{AGMOO00}
O.~Aharony, S.~S. Gubser, J.~Maldacena, H.~Ooguri, and Y.~Oz, { Large ${N}$
  field theories, string theory and gravity},  { Phys. Rept.} { 323} (2000)
  183--386,
[\href{http://xxx.lanl.gov/abs/hep-th/9905111}{{\tt hep-th/9905111}}].
%%CITATION = PRPLC,323,183;%%.

\bibitem{MalStr98}
J.~Maldacena and A.~Strominger, { {A}d{S}$_3$ black holes and a stringy
  exclusion principle},  { JHEP} { 12} (1998) 005,
[\href{http://xxx.lanl.gov/abs/hep-th/9804085}{{\tt hep-th/9804085}}].
%%CITATION = JHEPA,9812,005;%%.

\bibitem{Mart98}
E.~J. Martinec, { Matrix models of {A}d{S} gravity},
\href{http://xxx.lanl.gov/abs/hep-th/9804111}{{\tt hep-th/9804111}}.
%%CITATION = HEP-TH 9804111;%%.

\bibitem{GiKuSe98}
A.~Giveon, D.~Kutasov, and N.~Seiberg, { Comments on string theory on
  {A}d{S}$_3$},  { Adv. Theor. Math. Phys.} { 2} (1998) 733--780,
[\href{http://xxx.lanl.gov/abs/hep-th/9806194}{{\tt hep-th/9806194}}];
%%CITATION = 00203,2,733;%%.
D.~Kutasov and N.~Seiberg, { More comments on string theory on {A}d{S}$_3$},  {
  JHEP} { 04} (1999) 008,
[\href{http://xxx.lanl.gov/abs/hep-th/9903219}{{\tt hep-th/9903219}}].
%%CITATION = HEP-TH 9903219;%%.

\bibitem{SeiWit99}
N.~Seiberg and E.~Witten, { The {D}1/{D}5 system and singular {CFT}},  { JHEP}
  { 04} (1999) 017,
[\href{http://xxx.lanl.gov/abs/hep-th/9903224}{{\tt hep-th/9903224}}].
%%CITATION = HEP-TH 9903224;%%.

\bibitem{DaMaWa02}
J.~R. David, G.~Mandal, and S.~R. Wadia, { Microscopic formulation of black
  holes in string theory},  { Phys. Rept.} { 369} (2002) 549--686,
[\href{http://xxx.lanl.gov/abs/hep-th/0203048}{{\tt hep-th/0203048}}].
%%CITATION = HEP-TH 0203048;%%.

\bibitem{Giveon:2003ku}
A.~Giveon and A.~Pakman,
More on superstrings in {A}d{S}$_3 \times$ N,
JHEP {0303}, (2003) 056,
[\href{http://xxx.lanl.gov/abs/hep-th/0302217}{{\tt hep-th/0302217}}].
%%CITATION = HEP-TH 0302217;%%

\bibitem{Vafa95}
C.~Vafa, { Instantons on {D}-branes},  { Nucl. Phys.} { B463} (1996) 435--442,
[\href{http://xxx.lanl.gov/abs/hep-th/9512078}{{\tt hep-th/9512078}}].
%%CITATION = HEP-TH 9512078;%%.

\bibitem{Witt97}
E.~Witten, { On the conformal field theory of the {H}iggs branch},  { JHEP} {
  07} (1997) 003,
[\href{http://xxx.lanl.gov/abs/hep-th/9707093}{{\tt hep-th/9707093}}].
%%CITATION = HEP-TH 9707093;%%.

\bibitem{LarMar99}
F.~Larsen and E.~J. Martinec, { ${U}(1)$ charges and moduli in the {D}1-{D}5
  system},  { JHEP} { 06} (1999) 019,
[\href{http://xxx.lanl.gov/abs/hep-th/9905064}{{\tt hep-th/9905064}}].
%%CITATION = HEP-TH 9905064;%%.

\bibitem{JeMiRa99}
A.~Jevicki, M.~Mihailescu, and S.~Ramgoolam, { Gravity from {CFT} on
  ${S}^{{N}}({X})$: Symmetries and interactions},  { Nucl. Phys.} { B577}
  (2000) 47--72,
[\href{http://xxx.lanl.gov/abs/hep-th/9907144}{{\tt hep-th/9907144}}].
%%CITATION = HEP-TH 9907144;%%.

\bibitem{LunMat01}
O.~Lunin and S.~D. Mathur, { Three-point functions for ${M^N/S_N}$ orbifolds
  with ${{\cal N}} = 4$ supersymmetry},  { Commun. Math. Phys.} { 227} (2002)
  385--419,
[\href{http://xxx.lanl.gov/abs/hep-th/0103169}{{\tt hep-th/0103169}}].
%%CITATION = HEP-TH 0103169;%%.

\bibitem{Lars98}
F.~Larsen, { The perturbation spectrum of black holes in ${N} = 8$
  supergravity},  { Nucl. Phys.} { B536} (1998) 258--278,
[\href{http://xxx.lanl.gov/abs/hep-th/9805208}{{\tt hep-th/9805208}}].
%%CITATION = HEP-TH 9805208;%%.

\bibitem{dBoe98a}
J.~de~Boer, { Six-dimensional supergravity on ${S}^3 \times$ {A}d{S}$_3$ and 2d
  conformal field theory},  { Nucl. Phys.} { B548} (1999) 139--166,
[\href{http://xxx.lanl.gov/abs/hep-th/9806104}{{\tt hep-th/9806104}}];
%%CITATION = NUPHA,B548,139;%%.
{ Large ${N}$ elliptic genus and {AdS}/{CFT} correspondence},  {
  JHEP} { 05} (1999) 017,
[\href{http://xxx.lanl.gov/abs/hep-th/9812240}{{\tt hep-th/9812240}}].
%%CITATION = HEP-TH 9812240;%%.

\bibitem{MaMoSt99}
J.~Maldacena, G.~W. Moore, and A.~Strominger, { Counting {B}{P}{S} black holes
  in toroidal type {I}{I} string theory},
\href{http://xxx.lanl.gov/abs/hep-th/9903163}{{\tt hep-th/9903163}}.
%%CITATION = HEP-TH 9903163;%%.

\bibitem{GHMN00}
E.~Gava, A.~B. Hammou, J.~F. Morales, and K.~S. Narain, { {D}1/{D}5 systems in
  ${N} = 4$ string theories},  { Nucl. Phys.} { B605} (2001) 17--63,
[\href{http://xxx.lanl.gov/abs/hep-th/0012118}{{\tt hep-th/0012118}}];
%%CITATION = HEP-TH 0012118;%%.
{ Ad{S}/{CFT}
  correspondence and {D}$1$/{D}$5$ systems in theories with $16$ supercharges},
   { JHEP} { 03} (2001) 035,
[\href{http://xxx.lanl.gov/abs/hep-th/0102043}{{\tt hep-th/0102043}}].
%%CITATION = HEP-TH 0102043;%%.

\bibitem{FMMR98}
D.~Z. Freedman, S.~D. Mathur, A.~Matusis, and L.~Rastelli, { Correlation
  functions in the {CFT}$_d$/{A}d{S}$_{d+1}$ correspondence},  { Nucl. Phys.} {
  B546} (1999) 96--118,
[\href{http://xxx.lanl.gov/abs/hep-th/9804058}{{\tt hep-th/9804058}}].
%%CITATION = HEP-TH 9804058;%%.

\bibitem{LMRS98}
S.-M. Lee, S.~Minwalla, M.~Rangamani, and N.~Seiberg, { Three-point functions
  of chiral operators in ${D} = 4, {{\cal N}} = 4$ {SYM} at large ${N}$},  {
  Adv. Theor. Math. Phys.} { 2} (1998) 697--718,
[\href{http://xxx.lanl.gov/abs/hep-th/9806074}{{\tt hep-th/9806074}}].
%%CITATION = HEP-TH 9806074;%%.

\bibitem{Lee99}
S.-M. Lee, { Ad{S}$_5$/{CFT}$_4$ four-point functions of chiral primary
  operators: cubic vertices},  { Nucl. Phys.} { B563} (1999) 349--360,
[\href{http://xxx.lanl.gov/abs/hep-th/9907108}{{\tt hep-th/9907108}}].
%%CITATION = HEP-TH 9907108;%%.

\bibitem{AruFro99}
G.~Arutyunov and S.~Frolov, { Some cubic couplings in type {IIB} supergravity
  on {AdS}$_5 \times {S}^5$ and three-point functions in four-dimensional super
  {Y}ang-{M}ills theory at large~${N}$},  { Phys. Rev.} { D61} (2000) 064009,
[\href{http://xxx.lanl.gov/abs/hep-th/9907085}{{\tt hep-th/9907085}}];
%%CITATION = HEP-TH 9907085;%%.
{ Four-point functions of lowest weight chiral
  primary operators in ${N} = 4$ four-dimensional supersymmetric {Y}ang-{M}ills
  theory in the supergravity approximation},  { Phys. Rev.} { D62} (2000)
  064016,
[\href{http://xxx.lanl.gov/abs/hep-th/0002170}{{\tt hep-th/0002170}}].
%%CITATION = HEP-TH 0002170;%%.

\bibitem{Miha99}
M.~Mihailescu, { Correlation functions for chiral primaries in ${D} = 6$
  supergravity on {AdS}$_3 \times {S}^3$},  { JHEP} { 02} (2000) 007,
[\href{http://xxx.lanl.gov/abs/hep-th/9910111}{{\tt hep-th/9910111}}].
%%CITATION = HEP-TH 9910111;%%.

\bibitem{ArPaTh00}
G.~Arutyunov, A.~Pankiewicz, and S.~Theisen, { Cubic couplings in ${D} = 6, {N}
  = 4b$ supergravity on {AdS}$_3 \times {S}^3$},  { Phys. Rev.} { D63} (2001)
  044024,
[\href{http://xxx.lanl.gov/abs/hep-th/0007061}{{\tt hep-th/0007061}}].
%%CITATION = HEP-TH 0007061;%%.

\bibitem{GuRoWa86}
M.~G{\"u}naydin, L.~J. Romans, and N.~P. Warner, { Compact and noncompact
  gauged supergravity theories in five-dimensions},  { Nucl. Phys.} { B272}
  (1986)
598--646.
%%CITATION = NUPHA,B272,598;%%.

\bibitem{PilWar00}
K.~Pilch and N.~P. Warner, { A new supersymmetric compactification of chiral
  {IIB} supergravity},  { Phys. Lett.} { B487} (2000) 22--29,
[\href{http://xxx.lanl.gov/abs/hep-th/0002192}{{\tt hep-th/0002192}}].
%%CITATION = HEP-TH 0002192;%%.

\bibitem{deWNic87}
B.~de~Wit and H.~Nicolai, { The consistency of the ${S}^7$ truncation in
  $d\!=\!11$ supergravity},  { Nucl. Phys.} { B281} (1987)
211.
%%CITATION = NUPHA,B281,211;%%.

\bibitem{NaVavN99}
H.~Nastase, D.~Vaman, and P.~van Nieuwenhuizen, { Consistent nonlinear {K}{K}
  reduction of $11d$ supergravity on ${A}d{S}_7\!\times\!{S}_4$ and
  self-duality in odd dimensions},  { Phys. Lett.} { B469} (1999) 96--102,
[\href{http://xxx.lanl.gov/abs/hep-th/9905075}{{\tt hep-th/9905075}}].
%%CITATION = PHLTA,B469,96;%%.

\bibitem{FGPW99}
D.~Z. Freedman, S.~S. Gubser, K.~Pilch, and N.~P. Warner, { Renormalization
  group flows from holography --- supersymmetry and a $c$-theorem},  { Adv.
  Theor. Math. Phys.} { 3} (1999)
[\href{http://xxx.lanl.gov/abs/hep-th/9904017}{{\tt hep-th/9904017}}].
%%CITATION = HEP-TH 9904017;%%.

\bibitem{GPPZ00}
L.~Girardello, M.~Petrini, M.~Porrati, and A.~Zaffaroni, { The supergravity
  dual of ${N}=1$ super {Y}ang-{M}ills theory},  { Nucl. Phys.} { B569} (2000)
  451--469,
[\href{http://xxx.lanl.gov/abs/hep-th/9909047}{{\tt hep-th/9909047}}].
%%CITATION = HEP-TH 9909047;%%.

\bibitem{BiFrSk01}
M.~Bianchi, D.~Z. Freedman, and K.~Skenderis, { How to go with an {R}{G} flow},
   { JHEP} { 08} (2001) 041,
[\href{http://xxx.lanl.gov/abs/hep-th/0105276}{{\tt hep-th/0105276}}].
%%CITATION = HEP-TH 0105276;%%.

\bibitem{DKSS98}
S.~Deger, A.~Kaya, E.~Sezgin, and P.~Sundell, { Spectrum of ${D}\!=\!6$,
  ${N}\!=\!4b$ supergravity on {A}d{S}$_3 \!\times\! {S}^3$},  { Nucl. Phys.} {
  B536} (1998) 110--140,
[\href{http://xxx.lanl.gov/abs/hep-th/9804166}{{\tt hep-th/9804166}}].
%%CITATION = NUPHA,B536,110;%%.

\bibitem{NicSam00}
H.~Nicolai and H.~Samtleben, { Maximal gauged supergravity in three
  dimensions},  { Phys. Rev. Lett.} { 86} (2001) 1686--1689,
[\href{http://xxx.lanl.gov/abs/hep-th/0010076}{{\tt hep-th/0010076}}];
%%CITATION = PRLTA,86,1686;%%.
{ Compact and noncompact gauged maximal
  supergravities in three dimensions},  { JHEP} { 0104} (2001) 022,
[\href{http://xxx.lanl.gov/abs/hep-th/0103032}{{\tt hep-th/0103032}}].
%%CITATION = HEP-TH 0103032;%%.

\bibitem{NicSam01b}
H.~Nicolai and H.~Samtleben, { ${N} = 8$ matter coupled {A}d{S}$_3$
  supergravities},  { Phys. Lett.} { B514} (2001) 165--172,
[\href{http://xxx.lanl.gov/abs/hep-th/0106153}{{\tt hep-th/0106153}}].
%%CITATION = HEP-TH 0106153;%%.

\bibitem{LuPoSe02}
H.~Lu, C.~N. Pope, and E.~Sezgin, { ${SU}(2)$ reduction of six-dimensional
  $(1,0)$ supergravity},
Nucl.\ Phys.\ B {668}, (2003) 237,
[\href{http://xxx.lanl.gov/abs/hep-th/0212323}{{\tt hep-th/0212323}}].
%%CITATION = HEP-TH 0212323;%%.

\bibitem{NicSam03a}
H.~Nicolai and H.~Samtleben, { Chern-{S}imons vs.\ {Y}ang-{M}ills gaugings in
  three dimensions},
Nucl.\ Phys.\ B {668} (2003) 167,
[\href{http://xxx.lanl.gov/abs/hep-th/0303213}{{\tt hep-th/0303213}}].
%%CITATION = HEP-TH 0303213;%%.

\bibitem{LuPoSe03}
H.~Lu, C.~N. Pope, and E.~Sezgin,
{ Yang-{M}ills-{C}hern-{S}imons
  supergravity},
Class.\ Quant.\ Grav.\  {\bf 21}, (2004) 2733,
[\href{http://xxx.lanl.gov/abs/hep-th/0305242}{{\tt hep-th/0305242}}].
%%CITATION = HEP-TH 0305242;%%.

\bibitem{dWHeSa03}
B.~de~Wit, I.~Herger, and H.~Samtleben, { Gauged locally
  supersymmetric ${D} =  3$ nonlinear sigma models},
{Nucl. Phys.} {B671} (2003) 175--216,
[\href{http://xxx.lanl.gov/abs/hep-th/0307006}{{\tt hep-th/0307006}}].
%%CITATION = HEP-TH 0307006;%%

\bibitem{FiNiSa03}
T.~Fischbacher, H.~Nicolai, and H.~Samtleben, 
{Non-semisimple and complex gaugings of $N=16$ supergravity},
Commun.\ Math.\ Phys.\  {249} (2004) 475,
[\href{http://xxx.lanl.gov/abs/hep-th/0306276}{{\tt hep-th/0306276}}].
%%CITATION = HEP-TH 0306276;%%

\bibitem{GivPor90}
A.~Giveon and M.~Porrati, { A completely duality invariant effective action of
  ${N}=4$ heterotic strings},  { Phys. Lett.} { B246} (1990)
54--60;
%%CITATION = PHLTA,B246,54;%%.
{ Duality invariant string algebra and ${D} = 4$
  effective actions},  { Nucl. Phys.} { B355} (1991)
422--454.
%%CITATION = NUPHA,B355,422;%%.

\bibitem{Roma86}
L.~J. Romans, { Selfduality for interacting fields: Covariant field equations
  for six-dimensional chiral supergravities},  { Nucl. Phys.} { B276} (1986)
71--92.
%%CITATION = NUPHA,B276,71;%%.

\bibitem{AchTow86}
A.~Ach{\'u}carro and P.~K. Townsend, { A {C}hern-{S}imons action for
  three-dimensional anti-de {S}itter supergravity theories},  { Phys. Lett.} {
  B180} (1986)
89--92.
%%CITATION = PHLTA,B180,89;%%.

\bibitem{Davi99}
J.~R. David, { Anti-de {S}itter gravity associated with the supergroup
  ${SU}(1,1|2) \times {SU}(1,1|2)$},  { Mod. Phys. Lett.} { A14} (1999)
  1143--1148,
[\href{http://xxx.lanl.gov/abs/hep-th/9904068}{{\tt hep-th/9904068}}].
%%CITATION = HEP-TH 9904068;%%.

\bibitem{BerSam01}
M.~Berg and H.~Samtleben, { An exact holographic {RG} flow between 2d conformal
  fixed points},  { JHEP} { 05} (2002) 006,
[\href{http://xxx.lanl.gov/abs/hep-th/0112154}{{\tt hep-th/0112154}}].
%%CITATION = HEP-TH 0112154;%%.

\bibitem{FiNiSa02}
T.~Fischbacher, H.~Nicolai, and H.~Samtleben, { Vacua of maximal gauged ${D} =
  3$ supergravities},  { Class. Quant. Grav.} { 19} (2002) 5297--5334,
[\href{http://xxx.lanl.gov/abs/hep-th/0207206}{{\tt hep-th/0207206}}].
%%CITATION = HEP-TH 0207206;%%.

\end{thebibliography}

\providecommand{\href}[2]{#2}\begingroup\raggedright\endgroup

}

\end{document}